\documentclass[english,nofootinbib]{revtex4-1}
\usepackage[T1]{fontenc}
\usepackage[latin9]{inputenc}
\usepackage{geometry}
\usepackage{color}
\usepackage{babel}
\usepackage{amsmath}
\usepackage{amssymb}
\usepackage{graphicx}
\usepackage[unicode=true,
 bookmarks=false,
 breaklinks=false,pdfborder={0 0 1},backref=section,colorlinks=false]
 {hyperref}
\usepackage{breakurl}

\makeatletter

\usepackage{wasysym}
\usepackage{amsfonts}
\usepackage{babel}
\providecommand{\U}[1]{\protect\rule{.1in}{.1in}}

\makeatother

\begin{document}

\title{Higher-order modified Starobinsky inflation}

\author{R.R. Cuzinatto$^{1,2}$}
\email{rodrigo.cuzinatto@unifal-mg.edu.br}

\selectlanguage{english}%

\author{L.G. Medeiros$^{3,4}$}
\email{leogmedeiros@ect.ufrn.br}

\selectlanguage{english}%

\author{P.J. Pompeia$^{5}$\bigskip{}
}
\email{pompeia@ita.br}

\selectlanguage{english}%

\affiliation{$^{1}$Instituto de Ci\^encia e Tecnologia, Universidade Federal de
Alfenas, Rodovia Jos\'e Aur\'elio Vilela, 11999, Cidade Universit\'aria,
CEP 37715-400 Po\c cos de Caldas, Minas Gerais, Brazil.\bigskip{}
}

\affiliation{$^{2}$Department of Physics, McGill University, Ernest Rutherford
Physics Building, 3600 University Street, H3A 2T8, Montreal, Quebec,
Canada.\bigskip{}
}

\affiliation{$^{3}$Instituto de F\'isica Te\'orica, Universidade Estadual Paulista,
Rua Bento Teobaldo Ferraz, 271, Bloco II, P.O. Box 70532-2, CEP 01156-970
S\~ao Paulo, S\~ao Paulo, Brazil.\bigskip{}
}

\affiliation{$^{4}$Escola de Ci\^encia e Tecnologia, Universidade Federal do Rio
Grande do Norte, Campus Universit\'ario, s/n-{}-Lagoa Nova, CEP 59078-970
Natal, Rio Grande do Norte, Brazil.\bigskip{}
}

\affiliation{$^{5}$Departamento de F\'isica, Instituto Tecnol\'ogico da Aeron\'autica,
Pra\c ca Mal. Eduardo Gomes, 50, CEP 12228-900 S\~ao Jos\'e dos Campos, S\~ao
Paulo, Brazil.\bigskip{}
}




\begin{abstract}
An extension of the Starobinsky model is proposed. Besides the usual
Starobinsky Lagrangian, a term proportional to the derivative of the
scalar curvature, $\nabla_{\mu}R\nabla^{\mu}R$, is considered. The
analyzis is done in the Einstein frame with the introduction of a
scalar field and a vector field. We show that inflation is attainable
in our model, allowing for a graceful exit. We also build the cosmological
perturbations and obtain the leading-order curvature power spectrum,
scalar and tensor tilts and tensor-to-scalar ratio. The tensor and
curvature power spectrums are compared to the most recent observations
from BICEP2/Keck collaboration. We verify that the scalar-to-tensor
rate $r$ can be expected to be up to three times the values predicted
by Starobinsky model.
\end{abstract}
\maketitle


\section{Introduction\label{sec:Introduction}}

As is well known, the Starobinsky model is currently
the most promising one for describing the cosmological inflation \cite{Starobinski1980}.
Historically, several approaches were proposed exploring different
physical aspects for describing the first stages of the universe \cite{Starobinski1980,Starobinski1979,Starobinsky1979a,Sato1981,Guth1981}.
The hypothesis of an inflationary universe driven
by an scalar field was proposed in 1981 \cite{Sato1981,Guth1981}
and provided an ingenious apparatus to solve, with only one mechanism,
three disturbing problems of the standard big bang cosmology, namely
the horizon, flatness and magnetic monopole problems. As a bonus,
the large-scale structure can also be explained in this scenario.
Although it is remarkable that this proposition could solve these
problems at once, other complications arose, like the absence of a
smooth transition from a de Sitter-like expansion to a decelerated
Friedmann-Lema\^itre-Robertson-Walker (FLRW) one (a shortcoming dubbed
as ``graceful exit problem\textquotedblright ).\footnote{The paper \cite{Aldrovandi2008} addresses this transition from a
purely geometrical standpoint.} This motivated the proposition of alternative models (for instance,
\cite{Linde1982,Linde1983,Steinhardt1982,Lucchin1985,Steinhardt1989})
which solved the graceful exit problem at the cost of imposing a fine-tuning
on the effective potential parameters \cite{Steinhardt1989,Steinhardt1993}.
So the search for a consistent inflation model continued to strive
for accomplishing some specific goals \cite{Steinhardt1993}: (i)
providing a mechanism to drive the universe through a phase transition
from a false vacuum to a true vacuum state, (ii) generating a brief
period of exponential-like growth for the scale factor, and (iii)
stipulating a smooth ending for the highly accelerated growth (graceful
exit) thus allowing for the universe to reheat and enter a period
of decelerated FLRW expansion.

A plethora of models were proposed suggesting the existence of a single
scalar field (the ``inflaton'') or multiple scalar fields \cite{Linde1994,Wand2008}
(see also \cite{Basset2006} and references therein) that would drive
the inflationary process. These models essentially consist of matter
fields evolving in a curved spacetime described by general relativity
(GR).

A different category of inflationary models, which is of particular
interest here, is composed of those assuming modifications on the
underlying theory of gravitation (i.e. GR). The $f\left(R\right)$
theories of gravity are perhaps the most explored class of modified
gravity theories in the literature. An important feature of $f\left(R\right)$
theories lies on the fact that they are proven to be equivalent to
scalar-tensor theories\footnote{This equivalence is completely established at the classical level.
At quantum level, this equivalence occurs in the case of on-shell
quantum corrections whereas it is broken off shell \cite{Ruf2018}.} \cite{Faraoni2010,NojiriOdintsov2011,Capozziello2011}. This is very
useful since the techniques developed for treating inflation models
with scalar fields are applicable for an $f\left(R\right)$ theory
when it is considered on its equivalent scalar-tensor form. It is
important to recall that the scalar-tensor theory can be analyzed
both in Jordan and Einstein frames. Although these frames are related
by a conformal transformation, the analyzis of scalar-tensor models
on non-minimal inflationary contexts may lead to different predictions
in each case \cite{Karam2017,Shokri2017}.

Several inflation models have been proposed in the context of $f\left(R\right)$
gravity \cite{Amin2016,Artymoski2016,Elizalde2011,Brooker2016,Odintsov2015,Sadeghi2015,Asaka2016,Nojiri:2017ncd,Fabris2017},
the most iconic one being the Starobinsky model \cite{Starobinski1980,Starobinski1979,Starobinsky1979a},
which modifies the gravitational Lagrangian by adding to the usual
Einstein-Hilbert Lagrangian a term proportional to the square of the
scalar curvature, $L=R+aR^{2}$. This inflationary model is characterized
by being simultaneously minimal in its new features and especially
favoured by the most recent data from Planck satellite \cite{Calmet2016}.
For instance, Starobinsky's model predicts a tensor-to-scalar ratio $r<0.0048$
for a number of \emph{e}-folds
greater than $50$ (in a very conservative estimation) while Planck
data \cite{Planck2018} suggests $r<0.064$ in the best scenario.
This is one example of how superbly compatible with experimental data
Starobinsky's model is. However, this difference in order of magnitude
in the estimation of $r$ (due to the not yet so precise measurement
of this parameter) and the one predicted by the Starobinsky model
still allows other models to be compatible with the data. In particular,
those models that do not predict a very low production of primordial
gravitational waves cannot be discarded until a
precision of order $10^{-3}$ for the estimation of $r$ is finally
achieved.

It is interesting to note that in the context of $f\left(R\right)$
Lagrangians terms proportional to $R^{n},\,n\geq3$ are apparently
suppressed \cite{Huang2014}. Hence, analytical functions of $R$
would only give contributions equivalent to Starobinky's. In order
to generalize the $f\left(R\right)$ models in the inflationary context
other categories of modified gravity theories are taken into account,
for instance those with Lagrangians containing the Gauss-Bonnet invariant
and/or the Weyl tensor \cite{Starobinsky1987,Clunan2009,Capozzielo2015,Odintsov2015a,Sebastiani2015,Ivanov2016,Salvio2017}.
Some applications considering both inflationary scalar field and modified
theory of gravity can also be found in the literature \cite{Starobinsky1985,Starobinsky1991,Weinberg2008,Baumann2016}
with some interesting results, e.g. vector fields contribution
should no longer be ignored in the presence of a (square) Weyl term
in the Lagrangian \cite{Deruelle2010}.

Another category of modified gravity is composed of theories with
Lagrangians containing derivatives of the curvature tensors (Riemann,
Ricci, scalar curvature and so forth), which lead to field equations
with derivatives of the metric of order higher than four. They are usually
motivated in the context of quantum gravity and can be separated in
two sub-categories: (i) theories with infinite derivatives of curvature
\cite{Biswas2012,Biswas2015,Biswas2017,Modesto2014,Modesto2017,Shapiro2015}
and (ii) theories with finite derivatives of curvature \cite{Asorey1992,Accioly2017,ModestoShapiro2016,Shapiro2014,Shapiro2014a,Decanini2007,NosEPJC2008}.
While the latter can exhibit (super-)renormalizability and locality, they
are usually plagued with ghosts; the former, on their turn, may be
ghost-free but present non-locality \cite{ModestoShapiro2016}. Applications
of both approaches to inflationary context are found in the literature
\cite{Berkin1990,Gottlober1990,Gottlober1991,Amendola1993,Iihoshi2011,Castellanos2018,Diamandis2017,Koshelev2016,Koshelev2018,Edholm2017,Chialva2015}.
In particular, theories with an infinite number of derivatives are
able to modify the tensor-to-scalar ratio \cite{Koshelev2016,Koshelev2018,Edholm2017}.

However, if one wants to keep locality, then theories with a finite
number of derivatives are in order. As has been shown in Ref. \cite{Wands1994},
one can substitute the extra degrees of freedom associated to higher
derivatives by auxiliary scalar fields in a particular class of finite
higher order theory. In the case of sixth-order derivative equations
for the metric,\footnote{Usually Starobinsky Lagrangian plus a higher derivative term.}
inflation is carried out by two scalar fields \cite{Berkin1990,Gottlober1991,Amendola1993,Iihoshi2011,Castellanos2018}.
A new approach to deal with higher derivative Lagrangians has been
proposed \cite{NosPRD2016} where the extra degrees of freedom arising
from higher order contribution are replaced by auxiliary tensor fields
instead of scalar fields only. This approach (in Jordan frame) is
applicable for the class of higher order theories that is regular
in the sense discussed in Ref. \cite{NosPRD2016}. That paper verified
that the Lagrangian $L=R+aR^{2}+bR\square R$ analyzed in Refs. \cite{Gottlober1990,Iihoshi2011}
can be equivalently described by a Lagrangian containing a scalar
field and a vector field (instead of two scalar fields). The study
of the field equation demonstrated that this vector field has only one
(unconstrained) degree of freedom, showing consistency between the
results in \cite{NosPRD2016} and \cite{Gottlober1990,Iihoshi2011,Castellanos2018}.

In the present work, an inflationary model constructed by the addition
a higher order term of the type $\nabla_{\mu}R\nabla^{\mu}R$ to the
Starobinsky action is proposed. The model is described in the Einstein
frame, within the framework presented in Ref. \cite{Nos2018}. Accordingly,
the extra degrees of freedom are given by a scalar and by a vector
field. In this context, the scalar field plays the role of the usual
Starobinsky inflation while the vector field produces corrections
to Starobinsky's inflation. The study of background dynamics is done
under conditions that allow for an inflationary attractor regime obeying slow-roll conditions. The perturbative analyzis up to
slow-roll regime leading order is also performed showing how the term
$\nabla_{\mu}R\nabla^{\mu}R$ changes the predictions of Starobinsky
inflation.

The paper is organized as follows: In Section \ref{sec - MGA}, we
propose the modified gravity action and obtain the field equation
in terms of the metric and the auxiliary fields. Next, in Section
\ref{sec:Background}, we study the background equations and show
that an inflationary regime is attainable in our model. Section \ref{sec:Perturbations}
is devoted to the analyzis of the perturbed cosmological equations,
whose solutions are evaluated in Section \ref{sec:Sol_pert_eq}. Finally,
the cosmological parameters are determined in Section \ref{sec - Const Cos Par}.
Section \ref{sec:Discussion} is dedicated to the discussions of the
main results.


\section{Modified gravity action\label{sec - MGA}}

Starobinsky gravity \cite{Starobinski1980}, described by the action
\begin{equation}
S_{\text{Sta}}=\frac{M_{Pl}^{2}}{2}\int d^{4}x\sqrt{-g}\left[R+\frac{1}{2\kappa_{0}}R^{2}\right],\label{Sta}
\end{equation}
emerges nowadays as the most promising model for the description of
the inflationary paradigm. Among the class of minimalist inflationary
models, i.e. those composed of a single parameter \cite{Martin2014},
Starobinsky inflation is the one that best fits the observations of
the CMB anisotropies \cite{Planck2018}. Besides, from a theoretical
point of view, this model has an excellent motivation since quadratic
terms involving the Riemann tensor arise naturally in a bottom-up
approach to the quantization of gravitation \cite{Stelle1977,Asorey1992,Biswas2012}.

For the reasons given in the previous section, it is reasonable to
expect that $S_{\text{Sta}}$ is not a fundamental action for gravity
in spite of the success of Starobinsky inflation. Therefore, corrections
to $S_{\text{Sta}}$ should exist. The first corrections to be considered
in a context of increasing energy scales are those of the same order
of $R^{2}$, i.e. those of the kind $R_{\mu\nu}R^{\mu\nu}$.\footnote{In principle, one could think of an extra term of the type $R_{\mu\nu\alpha\beta}R^{\mu\nu\alpha\beta}$.
However, this term can be absorbed in $R^{2}$ and $R_{\mu\nu}R^{\mu\nu}$
due to the existence of the Gauss-Bonnet topological invariant $G^{2}=R^{2}-4R_{\mu\nu}R^{\mu\nu}+R_{\mu\nu\alpha\beta}R^{\mu\nu\alpha\beta}$.} On the one hand, the addition of this term to the action (\ref{Sta})
makes gravitation a renormalizable theory; on the other hand, it introduces
ghosts, rendering the quantization process questionable \cite{Desser1974,Stelle1977}.

The next order of correction in action $S_{\text{Sta}}$ is composed
by terms of the type
\[
\left(R_{\ast\ast\ast\ast}\right)^{3}\text{ or }\left(\nabla^{\ast}R_{\ast\ast\ast\ast}\right)^{2},
\]
where $R_{\ast\ast\ast\ast}$ represents the Riemann tensor or any
of its contractions. Cubic terms of the type $\left(R_{\ast\ast\ast\ast}\right)^{3}$
are not essential for consistency in the standard quantization procedure
since they do not affect the structure of the propagator \cite{Asorey1992}.
Thus, for simplicity, we neglect the cubic corrections and take into
account only the terms involving derivatives of curvature-based objects.
Ref. \cite{NosEPJC2008} showed that there are only four distinct
terms of the form $\left(\nabla^{\ast}R_{\ast\ast\ast\ast}\right)^{2}$,
namely,
\[
\nabla_{\mu}R\nabla^{\mu}R\text{; \ }\nabla_{\mu}R_{\alpha\beta}\nabla^{\mu}R^{\alpha\beta}\text{; \ }\nabla^{\mu}R^{\alpha\beta}\nabla_{\alpha}R_{\mu\beta}\text{ \ and }\nabla_{\rho}R_{\mu\nu\alpha\beta}\nabla^{\rho}R^{\mu\nu\alpha\beta}\text{.}
\]
By using Bianchi identities, it is possible to verify that only two
of the four terms above are independent (modulo cubic order terms).
Therefore, the action integral with corrections to Einstein-Hilbert
term up to second order is:
\begin{equation}
S_{2}=\frac{M_{Pl}^{2}}{2}\int d^{4}x\sqrt{-g}\left[R+\frac{1}{2\kappa_{0}}R^{2}+\frac{1}{2\kappa_{2}}R_{\mu\nu}R^{\mu\nu}+\frac{\beta_{0}}{2\kappa_{0}^{2}}\nabla_{\mu}R\nabla^{\mu}R+\frac{\beta_{2}}{2\kappa_{2}^{2}}\nabla_{\mu}R_{\alpha\beta}\nabla^{\mu}R^{\alpha\beta}\right],\label{Acao com dois Riemanns}
\end{equation}
where $\kappa_{i}$ are constants with square mass dimension and $\beta_{i}$
are dimensionless constants. This action presents interesting properties
such as super-renormalizability \cite{Asorey1992,ModestoShapiro2016}
and finiteness of the gravitational potential (weak field regime)
at the origin \cite{AcciolyGiacchiniShapiro2016}. However, the presence
of the massive spin-$2$ terms associated with $R_{\mu\nu}$ inevitably
introduces ghosts into the theory.\footnote{In principle, non-local extensions of this action can make the theory
free of ghosts \cite{Biswas2012}.}

It may be conjectured that the pathologies associated with ghosts
(vacuum decay or unitarity loss \cite{Sbisa2014}) can be controlled
during the well-defined energy scales of inflation by making $S_{2}$
a consistent effective theory \cite{Fradkin1982,Buchbinder1989}.
This type of approach was adopted in Refs. \cite{Clunan2009,Ivanov2016,Salvio2017}
precisely to deal with inflationary models which contain the $R_{\mu\nu}R^{\mu\nu}$
term. Although this is a valid approach, in this work we will neglect
both the spin-$2$ terms.

Based on the previous discussion, we start by considering a gravitational
action that differs from $S_{\text{Sta}}$ by the addition of the
higher-order term $\nabla_{\mu}R\nabla^{\mu}R$:
\begin{equation}
S_{g}=\frac{M_{Pl}^{2}}{2}\int d^{4}x\sqrt{-g}\left[R+\frac{1}{2\kappa_{0}}R^{2}+\frac{\beta_{0}}{2\kappa_{0}^{2}}\nabla_{\mu}R\nabla^{\mu}R\right].\label{Sg}
\end{equation}
Constant $\kappa_{0}$ sets the energy scale of the inflationary regime
and $\beta_{0}$ is a measure of the deviation from Starobinsky inflation
model. An important point to be emphasized is that Eq. (\ref{Sg})
will be ghost-free if $\beta_{0}<0$ \cite{Hindawi1996}.\footnote{The metric signature adopted here is $\left(-,+,+,+\right)$.}
Ref. \cite{Nos2018} has shown this action can be re-expressed in
Einstein frame where a scalar field and a vector field play the role
of the higher derivative terms:
\begin{equation}
S_{g}^{\prime\prime}=\int d^{4}x\sqrt{-\tilde{g}}\left\{ \frac{M_{Pl}^{2}}{2}\tilde{R}-\frac{1}{2}\partial_{\rho}\tilde{\Phi}\tilde{\partial}^{\rho}\tilde{\Phi}-\frac{M_{Pl}^{2}}{2}e^{-2\sqrt{\frac{2}{3}}\frac{\tilde{\Phi}}{M_{Pl}}}\left(\frac{\kappa_{0}}{2}\Upsilon^{2}+\frac{1}{2}\frac{\beta_{0}}{\kappa_{0}^{2}}e^{-\sqrt{\frac{2}{3}}\frac{\tilde{\Phi}}{M_{Pl}}}\tilde{g}_{\mu\nu}\xi^{\nu}\xi^{\mu}\right)\right\} ,\label{Eq acao sem redefinicao}
\end{equation}
where
\[
\Upsilon\equiv\Upsilon\left(\tilde{\Phi},\partial\tilde{\Phi},\xi^{\mu},\partial\xi^{\mu}\right)\equiv e^{\sqrt{\frac{2}{3}}\frac{\tilde{\Phi}}{M_{Pl}}}+\frac{\beta_{0}}{\kappa_{0}^{2}}\tilde{\nabla}_{\mu}\xi^{\mu}-\frac{2}{M_{Pl}}\sqrt{\frac{2}{3}}\frac{\beta_{0}}{\kappa_{0}^{2}}\xi^{\rho}\partial_{\rho}\tilde{\Phi}-1,
\]
with
\begin{align*}
\tilde{\Phi} & \equiv M_{Pl}\sqrt{\frac{3}{2}}\ln\left(\frac{\partial f\left(\xi,\xi_{\mu}\right)}{\partial\xi}-\nabla_{\mu}\phi^{\mu}\right),\\
\phi^{\mu} & \equiv\frac{\partial f}{\partial\xi_{\mu}}=\frac{\beta_{0}}{\kappa_{0}^{2}}\xi^{\mu}.
\end{align*}
The effective ``matter\textquotedblright{} field Lagrangian, i.e.
the Lagrangian for the scalar and vector fields, now reads:
\begin{equation}
\mathcal{L}_{\text{eff}}=-\frac{1}{2}\partial_{\rho}\tilde{\Phi}\tilde{\partial}^{\rho}\tilde{\Phi}-\frac{M_{Pl}^{2}}{2}e^{-2\sqrt{\frac{2}{3}}\frac{\tilde{\Phi}}{M_{Pl}}}\left(\frac{\kappa_{0}}{2}\Upsilon^{2}+\frac{1}{2}\frac{\beta_{0}}{\kappa_{0}^{2}}e^{-\sqrt{\frac{2}{3}}\frac{\tilde{\Phi}}{M_{Pl}}}\tilde{g}_{\mu\nu}\xi^{\nu}\xi^{\mu}\right).\label{L_eff}
\end{equation}

Lagrangian $\mathcal{L}_{\text{eff}}$ is used for evaluating the
field equations for $\tilde{\Phi}$ and $\xi^{\mu}$, which are given
respectively by
\begin{gather}
\tilde{\Box}\tilde{\Phi}+\sqrt{\frac{2}{3}}\frac{\kappa_{0}}{2}M_{Pl}e^{-2\sqrt{\frac{2}{3}}\frac{\tilde{\Phi}}{M_{Pl}}}\Upsilon\left(\Upsilon-e^{\sqrt{\frac{2}{3}}\frac{\tilde{\Phi}}{M_{Pl}}}\right)\nonumber \\
+\frac{\beta_{0}}{\kappa_{0}^{2}}M_{Pl}\left[-\kappa_{0}\sqrt{\frac{2}{3}}\tilde{\nabla}_{\rho}\left(e^{-2\sqrt{\frac{2}{3}}\frac{\tilde{\Phi}}{M_{Pl}}}\Upsilon\xi^{\rho}\right)+\sqrt{\frac{2}{3}}\frac{3}{4}e^{-3\sqrt{\frac{2}{3}}\frac{\tilde{\Phi}}{M_{Pl}}}\tilde{\xi}_{\mu}\xi^{\mu}\right]=0,\label{Eq Phi sem def 1}
\end{gather}
and
\begin{equation}
\kappa_{0}\partial_{\rho}\Upsilon-e^{-\sqrt{\frac{2}{3}}\frac{\tilde{\Phi}}{M_{Pl}}}\tilde{\xi}_{\rho}=0,\label{Eq Psi sem def 1}
\end{equation}
where $\tilde{\xi}_{\mu}\equiv\tilde{g}_{\mu\nu}\xi^{\nu}$. These
two equations can be combined to give:
\begin{equation}
\tilde{\Box}\tilde{\Phi}+\sqrt{\frac{2}{3}}\frac{\kappa_{0}}{2}M_{Pl}e^{-2\sqrt{\frac{2}{3}}\frac{\tilde{\Phi}}{M_{Pl}}}\Upsilon\left(e^{\sqrt{\frac{2}{3}}\frac{\tilde{\Phi}}{M_{Pl}}}-\Upsilon-2\right)-\frac{\beta_{0}}{\kappa_{0}^{2}}M_{Pl}\sqrt{\frac{2}{3}}\frac{1}{4}e^{-3\sqrt{\frac{2}{3}}\frac{\tilde{\Phi}}{M_{Pl}}}\tilde{\xi}_{\mu}\xi^{\mu}=0.\label{Eq Phi sem def}
\end{equation}

The field equation for $\tilde{g}^{\mu\nu}$ reads:
\[
\tilde{G}_{\mu\nu}=\frac{1}{M_{Pl}^{2}}\tilde{T}_{\mu\nu}^{\left(\text{eff}\right)},
\]
where
\begin{align}
\tilde{T}_{\rho\sigma}^{\left(\text{eff}\right)} & \equiv\frac{2}{\sqrt{-\tilde{g}}}\frac{\delta\left(\sqrt{-\tilde{g}}\mathcal{L}_{\text{eff}}\right)}{\delta\tilde{g}^{\rho\sigma}}\nonumber \\
 & =\partial_{\rho}\tilde{\Phi}\partial_{\sigma}\tilde{\Phi}-\frac{M_{Pl}^{2}}{2}\frac{\beta_{0}}{\kappa_{0}^{2}}e^{-3\sqrt{\frac{2}{3}}\frac{\tilde{\Phi}}{M_{Pl}}}\tilde{\xi}_{\rho}\tilde{\xi}_{\sigma}+\tilde{g}_{\rho\sigma}\left[\mathcal{L}_{\text{eff}}+\frac{M_{Pl}^{2}}{2}\frac{\beta_{0}}{\kappa_{0}}\tilde{\nabla}_{\nu}\left(e^{-2\sqrt{\frac{2}{3}}\frac{\tilde{\Phi}}{M_{Pl}}}\Upsilon\xi^{\nu}\right)\right]\label{Tmunu sem definicao}
\end{align}
is the (effective) energy-momentum tensor.

Henceforth we shall omit the tilde for notation economy.

The effective energy-momentum tensor (\ref{Tmunu sem definicao})
is of the imperfect fluid type \cite{Nos2018}:
\begin{equation}
T_{\mu\nu}=\left(\varepsilon+p\right)u_{\mu}u_{\nu}+pg_{\mu\nu}+u_{\mu}q_{\nu}+u_{\nu}q_{\mu}+\pi_{\mu\nu},\label{Tmunu imperfect}
\end{equation}
where $p$ is the pressure, $\varepsilon$ is the energy density,
$u_{\mu}$ is the four-velocity of the fluid element, $q_{\mu}$ is
the heat flux vector and $\pi_{\mu\nu}$ is the viscous shear tensor;
these quantities satisfy $q_{\mu}u^{\mu}=0,\,\pi_{\mu\nu}u^{\nu}=0,\,\pi_{\,\mu}^{\mu}=0,\,\pi_{\mu\nu}=\pi_{\nu\mu}$.
In fact, Eqs. (\ref{Tmunu sem definicao}) and (\ref{Tmunu imperfect})
are the same under the following identifications:
\begin{align}
\varepsilon+p & =-\partial^{\alpha}\Phi\partial_{\alpha}\Phi+\frac{M_{Pl}^{2}}{2}\frac{\beta_{0}}{\kappa_{0}^{2}}e^{-3\sqrt{\frac{2}{3}}\frac{\Phi}{M_{Pl}}}\xi_{\alpha}\xi^{\alpha},\label{epsolon mais p FI}\\
u_{\mu} & =\frac{1}{N}\left(\partial_{\mu}\Phi+\sqrt{\frac{M_{Pl}^{2}}{2}\frac{\beta_{0}}{\kappa_{0}^{2}}}e^{-\frac{3}{2}\sqrt{\frac{2}{3}}\frac{\Phi}{M_{Pl}}}\xi_{\mu}\right),\label{4 velocidade FI}\\
q_{\mu} & =-\sqrt{\frac{M_{Pl}^{2}}{2}\frac{\beta_{0}}{\kappa_{0}^{2}}}e^{-\frac{3}{2}\sqrt{\frac{2}{3}}\frac{\Phi}{M_{Pl}}}N\left(\xi_{\mu}+\xi^{\alpha}u_{\alpha}u_{\mu}\right),\label{fluxo de calor FI}\\
p & =\mathcal{L}_{\text{eff}}+\frac{M_{Pl}^{2}}{2}\frac{\beta_{0}}{\kappa_{0}}\nabla_{\nu}\left(e^{-2\sqrt{\frac{2}{3}}\frac{\Phi}{M_{Pl}}}\Upsilon\xi^{\nu}\right),\label{pressao FI}\\
\pi_{\mu\nu} & =0,\label{Viscosidade}
\end{align}
with
\begin{equation}
N=\sqrt{-\left(\partial_{\alpha}\Phi+\sqrt{\frac{M_{Pl}^{2}}{2}\frac{\beta_{0}}{\kappa_{0}^{2}}}e^{-\frac{3}{2}\sqrt{\frac{2}{3}}\frac{\Phi}{M_{Pl}}}\xi_{\alpha}\right)\left(\partial^{\alpha}\Phi+\sqrt{\frac{M_{Pl}^{2}}{2}\frac{\beta_{0}}{\kappa_{0}^{2}}}e^{-\frac{3}{2}\sqrt{\frac{2}{3}}\frac{\Phi}{M_{Pl}}}\xi^{\alpha}\right)},\label{Normalization}
\end{equation}
Hence, the fluid represented by Eq. (\ref{Tmunu sem definicao}) has
no contribution from viscous shear components, which are null here.
Notice that the heat flux vector exists solely due to the higher order
term \textemdash{} were it absent, the theory would be reduced to
Starobinsky's model and, therefore, would be represented by a perfect
fluid energy-momentum tensor. The above equations are specified in
FLRW spacetime in the next section.


\section{Cosmological background equations\label{sec:Background}}

In order to analyze the action (\ref{Eq acao sem redefinicao}) for
background cosmology, we consider: (i) a homogeneous and isotropic
spacetime, (ii) a comoving reference frame ($u^{\mu}=\delta_{0}^{\mu}$),
(iii) spherical coordinates for the space sector and (iv) null space
curvature parameter ($k=0$). With these assumptions, the line element
is FLRW metric:
\[
ds^{2}=-dt^{2}+a^{2}\left(t\right)\left[dr^{2}+r^{2}d\Omega^{2}\right],
\]
where $a\left(t\right)$ is the scale factor.

In this case, the Einstein tensor is diagonal and Einstein equations
imply that the space components of the heat flux vector are null.
Also, condition $q_{\mu}u^{\mu}=0$ imposes $q_{0}=0$ showing there
is no heat flux for a homogeneous and isotropic spacetime. Einstein
equations are then reduced to:
\begin{align}
H^{2} & =\frac{1}{3M_{Pl}^{2}}\varepsilon,\label{Friedmann one1}\\
\frac{dH}{dt} & =-\frac{1}{2M_{Pl}^{2}}\left(\varepsilon+p\right),\label{Friedmann two2}
\end{align}
where $H\equiv\frac{1}{a}\frac{da}{dt}$ is the Hubble function. The
energy density and pressure are given in terms of the auxiliary fields
and their derivatives:
\begin{align*}
\varepsilon & =\frac{1}{2}\left(\frac{d\Phi}{dt}\right)^{2}+\frac{1}{2}\frac{M_{Pl}^{2}}{2}\kappa_{0}\left(1-e^{-\sqrt{\frac{2}{3}}\frac{\Phi}{M_{Pl}}}\right)^{2}-\frac{1}{2}\frac{M_{Pl}^{2}}{2}\frac{\beta_{0}}{\kappa_{0}^{2}}e^{-3\sqrt{\frac{2}{3}}\frac{\Phi}{M_{Pl}}}\left(\xi^{0}\right)^{2}\\
 & -\frac{1}{2}\frac{M_{Pl}^{2}}{2}\kappa_{0}\left(\frac{\beta_{0}}{\kappa_{0}^{2}}\right)^{2}e^{-2\sqrt{\frac{2}{3}}\frac{\Phi}{M_{Pl}}}\left(\frac{d\xi^{0}}{dt}+3H\xi^{0}-\frac{2}{M_{Pl}}\sqrt{\frac{2}{3}}\xi^{0}\frac{d\Phi}{dt}\right)^{2},
\end{align*}
and
\begin{equation}
\varepsilon+p=\left(\frac{d\Phi}{dt}\right)^{2}-\frac{M_{Pl}^{2}}{2}\frac{\beta_{0}}{\kappa_{0}^{2}}e^{-3\sqrt{\frac{2}{3}}\frac{\Phi}{M_{Pl}}}\left(\xi^{0}\right)^{2}.\label{energy plus pressure}
\end{equation}

For consistency, the auxiliary field $\Phi$ must also be homogenous
and isotropic, so it can only be time-dependent. In this case, when
the comoving frame is considered, the auxiliary vector field space
components have to be null, which is consistent with the fact that
the heat flux vector is also null. Notice that $\xi^{0}$ does not
vanish and actually it is a dynamical quantity in FLRW background,
as verified by the field equations (\ref{Eq Phi sem def}) and (\ref{Eq Psi sem def 1}):
\begin{equation}
\frac{d^{2}\Phi}{dt^{2}}+3H\frac{d\Phi}{dt}-\sqrt{\frac{2}{3}}\frac{\kappa_{0}}{2}M_{Pl}e^{-2\sqrt{\frac{2}{3}}\frac{\Phi}{M_{Pl}}}\Upsilon\left(e^{\sqrt{\frac{2}{3}}\frac{\Phi}{M_{Pl}}}-\Upsilon-2\right)-\frac{1}{4}\sqrt{\frac{2}{3}}\frac{\beta_{0}}{\kappa_{0}^{2}}M_{Pl}e^{-3\sqrt{\frac{2}{3}}\frac{\Phi}{M_{Pl}}}\left(\xi^{0}\right)^{2}=0,\label{Aux field eq 1}
\end{equation}
and
\begin{equation}
\kappa_{0}\frac{d\Upsilon}{dt}+e^{-\sqrt{\frac{2}{3}}\frac{\tilde{\Phi}}{M_{Pl}}}\xi^{0}=0,\label{Aux field eq 2}
\end{equation}
where
\[
\Upsilon=e^{\sqrt{\frac{2}{3}}\frac{\Phi}{M_{Pl}}}-1+\frac{\beta_{0}}{\kappa_{0}^{2}}\left(\frac{d\xi^{0}}{dt}+3H\xi^{0}-\frac{2}{M_{Pl}}\sqrt{\frac{2}{3}}\xi^{0}\frac{d\Phi}{dt}\right).
\]

The following step is to check whether the above equations are suitable
to describe an inflationary regime. This is the analysis in the next
subsection.


\subsection{Analysis of the field equations: attractors}

We work in phase space. It is convenient to define new dimensionless
variables for the auxiliary fields:
\begin{equation}
X\equiv\sqrt{\frac{2}{3}}\frac{\Phi}{M_{Pl}}\text{, \ \ }Y\equiv\sqrt{\frac{2}{3}}\frac{\dot{\Phi}}{M_{Pl}}\text{, }\quad T\equiv\frac{\xi^{0}}{\kappa_{0}^{3/2}}\text{ \ \ and \ \ \ }S\equiv\frac{\dot{\xi}^{0}}{\kappa_{0}^{3/2}},\label{Var adm}
\end{equation}
with
\begin{equation}
\bar{\partial}_{0}Q=\dot{Q}\equiv\frac{1}{\sqrt{\kappa_{0}}}\partial_{0}Q,\label{def 2}
\end{equation}
where $\bar{\partial}_{0}$ (and dot) denotes a dimensionless time
derivative. A dimensionless Hubble function can also be defined:
\begin{equation}
h_{\kappa}\equiv\frac{H}{\sqrt{\kappa_{0}}}.\label{def 3}
\end{equation}
Then, FLRW equations, (\ref{Friedmann one1}) and (\ref{Friedmann two2}),
become:
\begin{align}
\dot{h}_{\kappa} & =-\frac{1}{4}\left(3Y^{2}-\beta_{0}T^{2}e^{-3X}\right),\label{Hpto adm1}\\
h_{k}^{2} & =\frac{1}{12}\left[3Y^{2}+\left(1-e^{-X}\right)^{2}-\beta_{0}e^{-3X}T^{2}-\beta_{0}^{2}e^{-2X}\left(S+3h_{\kappa}T-2TY\right)^{2}\right].\label{H2 adm1}
\end{align}
Similarly, the auxiliary field equations (\ref{Aux field eq 1}) and
(\ref{Aux field eq 2}) are rewritten as :
\begin{align}
 & \dot{Y}+3h_{\kappa}Y+\frac{1}{3}e^{-X}\left(1-e^{-X}\right)-\frac{1}{6}\beta_{0}e^{-3X}T^{2}\nonumber \\
 & +\frac{1}{3}\beta_{0}e^{-X}\left(S+3h_{\kappa}T-2YT\right)+\frac{1}{3}\beta_{0}^{2}e^{-2X}\left(S+3h_{\kappa}T-2YT\right)^{2}=0,\label{Phi adm1}
\end{align}
and
\begin{equation}
Ye^{X}+\beta_{0}\left(\dot{S}+3\dot{h}_{\kappa}T+3h_{\kappa}S-2SY-2\dot{Y}T\right)+e^{-X}T=0.\label{Quisi adm1}
\end{equation}

The quadratic equation (\ref{H2 adm1}) can be manipulated to express
$h_{\kappa}$ in terms of the auxiliary fields $X,\,Y,\,T$ and $S$:
\begin{align}
h_{\kappa} & =\frac{-\frac{1}{2}\beta_{0}^{2}e^{-2X}T\left(S-2TY\right)}{2\left(1+\frac{3}{4}\beta_{0}^{2}e^{-2X}T^{2}\right)}\nonumber \\
 & +\frac{\sqrt{\left(1+\frac{3}{4}\beta_{0}^{2}e^{-2X}T^{2}\right)\frac{1}{3}\left[3Y^{2}+\left(1-e^{-X}\right)^{2}-\beta_{0}e^{-3X}T^{2}\right]-\frac{1}{3}\beta_{0}^{2}e^{-2X}\left(S-2TY\right)^{2}}}{2\left(1+\frac{3}{4}\beta_{0}^{2}e^{-2X}T^{2}\right)},\label{h adm1}
\end{align}
The positive sign in front of the square root must be chosen to recover
Staronbinsky's results in the limit $\beta_{0}\rightarrow0$. In addition,
there are two terms within the square root with negative signs; they
could eventually turn $h_{\kappa}$ into a complex number. As this
is meaningless in the present context, the phase space is constrained
to satisfy the condition:
\[
\left(1+\frac{3}{4}\beta_{0}^{2}e^{-2X}T^{2}\right)\frac{1}{3}\left[3Y^{2}+\left(1-e^{-X}\right)^{2}\right]\geq\beta_{0}e^{-3X}T^{2}\left(1+\frac{3}{4}\beta_{0}^{2}e^{-2X}T^{2}\right)+\frac{1}{3}\beta_{0}^{2}e^{-2X}\left(S-2TY\right)^{2}.
\]

Eq. (\ref{h adm1}) for $h_{\kappa}$ can be used in Eqs. (\ref{Phi adm1})
and (\ref{Quisi adm1}), so that an autonomous system is obtained:
\begin{equation}
\begin{cases}
\dot{X}=Y\\
\dot{Y}=j_{1}\left(X,Y,T,S\right)\\
\dot{T}=S\\
\beta_{0}\dot{S}=j_{2}\left(X,Y,T,S\right)
\end{cases}\label{AutSyst}
\end{equation}
where
\begin{align}
j_{1}\left(X,Y,T,S\right) & \equiv-3h_{\kappa}Y-\frac{1}{3}e^{-X}\left(1-e^{-X}\right)+\frac{1}{6}\beta_{0}e^{-3X}T^{2}\nonumber \\
 & -\frac{1}{3}\beta_{0}e^{-X}\left(S+3h_{\kappa}T-2YT\right)-\frac{1}{3}\beta_{0}^{2}e^{-2X}\left(S+3h_{\kappa}T-2YT\right)^{2},\label{j1}\\
j_{2}\left(X,Y,T,S\right) & =-Ye^{X}-e^{-X}T-3\beta_{0}h_{\kappa}S+2\beta_{0}SY\nonumber \\
 & -\frac{2}{3}\beta_{0}T\left[9h_{\kappa}Y+e^{-X}\left(1-e^{-X}\right)-\frac{1}{8}\left(27Y^{2}-\beta_{0}T^{2}e^{-3X}\right)\right.\nonumber \\
 & \left.+\beta_{0}e^{-X}\left(S+3h_{\kappa}T-2YT\right)+\beta_{0}^{2}e^{-2X}\left(S+3h_{\kappa}T-2YT\right)^{2}\right].\label{j2}
\end{align}
The dynamical system above characterizes higher-order modified Starobinsky
inflation on the background.


\subsubsection{Slow-roll regime and the end of inflation \label{InflationTermination}}

First, it is important to realize that $X=Y=T=S=0$ is a fixed point
of the phase space. If we are supposed to have an inflationary expansion
that endures for a certain finite period of time, this fixed point
has to be stable, i.e. trajectories in the phase space must tend
to the origin. The stability of this point can be determined by the
Lyapunov coefficients $\lambda$, which are the eigenvalues of the
linearization matrix $M$. The matrix entries are calculated as partial
derivatives of the right hand side of Eq. (\ref{AutSyst}) with respect
to $X,\,Y,\,T,\,S$:
\[
\left(\begin{array}{c}
\dot{X}\\
\dot{Y}\\
\dot{T}\\
\dot{S}
\end{array}\right)=M\left(\begin{array}{c}
X\\
Y\\
T\\
S
\end{array}\right),\qquad M=\left(\begin{array}{cccc}
0 & 1 & 0 & 0\\
-\frac{1}{3} & 0 & 0 & -\frac{\beta_{0}}{3}\\
0 & 0 & 0 & 1\\
0 & -\frac{1}{\beta_{0}} & -\frac{1}{\beta_{0}} & 0
\end{array}\right).
\]
The four eigenvalues are:
\[
\lambda_{mn}=\left(-1\right)^{m}\sqrt{\frac{1}{2\beta_{0}}\left(-1+\left(-1\right)^{n}\sqrt{\left(1-\frac{4}{3}\beta_{0}\right)}\right)}\qquad(m,n=1,2).
\]
It is clear the stability of the fixed point depends on the $\beta_{0}$
values.

We start by considering $\beta_{0}<0$. In this case,
\begin{align*}
\lambda_{21}= & \sqrt{\frac{1}{2\left|\beta_{0}\right|}+\frac{\sqrt{3}}{6\left|\beta_{0}\right|}\sqrt{\left(3+4\left|\beta_{0}\right|\right)}}>0,
\end{align*}
which implies the instability of the fixed point.

If we take $\beta_{0}>\frac{3}{4}$, then
\[
\lambda_{mn}=\left(-1\right)^{m}\sqrt{\frac{1}{2\beta_{0}}}\sqrt{-1+i\left(-1\right)^{n}\sqrt{\left(\frac{4}{3}\beta_{0}-1\right)}}.
\]
The coefficient is then a square root of a complex number, which splits
in a real part and a complex piece. This implies that at least one
of the eigenvalues will have a positive real part, leading to an instability
of the fixed point.

At last, in the interval $0<\beta_{0}<\frac{3}{4}$, we have $\sqrt{\left(1-\frac{4}{3}\beta_{0}\right)}<1$
and the Lyapunov coefficients $\lambda_{mn}$ become pure imaginary
numbers. Consequently, the fixed point is a center; the neighbouring
trajectories will remain convergent to this point.\footnote{An analogous situation occurs in the Starobinsky inflation ($\beta_{0}=0$).}
The physical consequence of this result is: the values of $\beta_{0}$
within the interval $\left[0,3/4\right]$ make it possible for inflation
to cease smoothly, allowing for reheating.


\begin{figure}[h]
\includegraphics[scale=0.278]{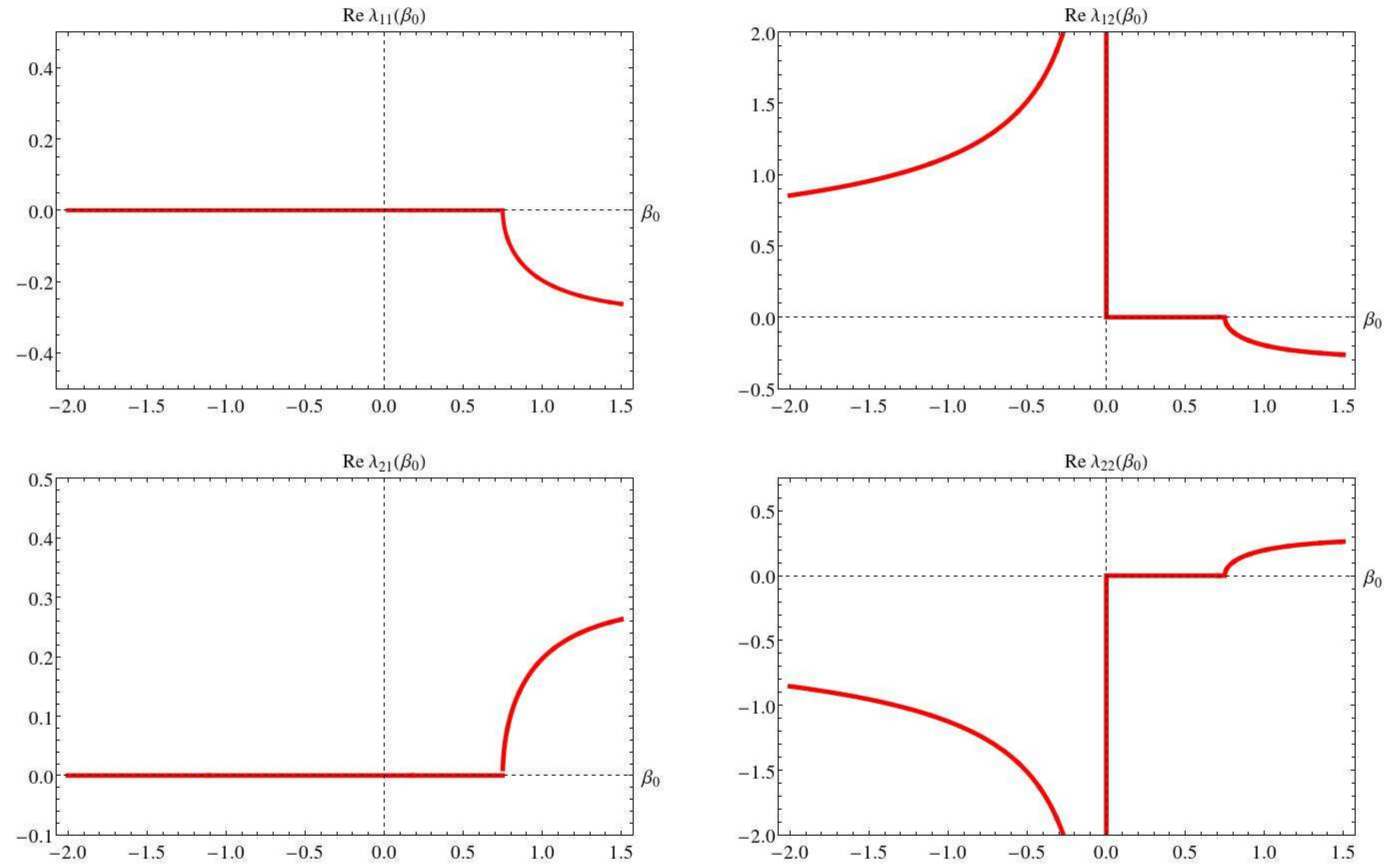}

\caption{Plots of the real parts of the Lyapunov coefficients as a function
of $\beta_{0}$.\emph{ \label{Figure_lambdas}}}
\end{figure}


The same conclusions can be obtained numerically. Fig. \ref{Figure_lambdas}
shows the real part of each $\lambda_{mn}$ plotted as a function
of $\beta_{0}$. The graphs show the existence of at least one eigenvalue
with a positive real part when $\beta_{0}>\frac{3}{4}$ or $\beta_{0}<0$
. For $0<\beta_{0}<\frac{3}{4}$, all the eigenvalues are pure imaginary
numbers.

At this point, it is interesting to recall some results presented
in Ref. \cite{Berkin1990}, where the authors consider a similar higher order
term in a double inflation scenario
(i.e. inflation from two scalar fields). They claim that in order to have ``a
large range of initial conditions\textquotedblright ,it should be $-\gamma\ll\alpha^{2}$. In our case, this condition
is equivalent to impose $\beta_{0}\ll1$. Moreover, the results above
suggest that $0<\beta_{0}<\frac{3}{4}$ is a necessary condition for
an inflationary scenario. The value $\frac{3}{4}$ is just an upper
limit below which inflation is attainable in our model. We still have
to analyze the existence of a slow-roll regime leading to a ``graceful
exit\textquotedblright{} (end of inflation). In what follows, we show
the value $3/4$ is still an overestimated upper limit for $\beta_{0}$.

\begin{figure}[h]
\includegraphics[scale=0.42]{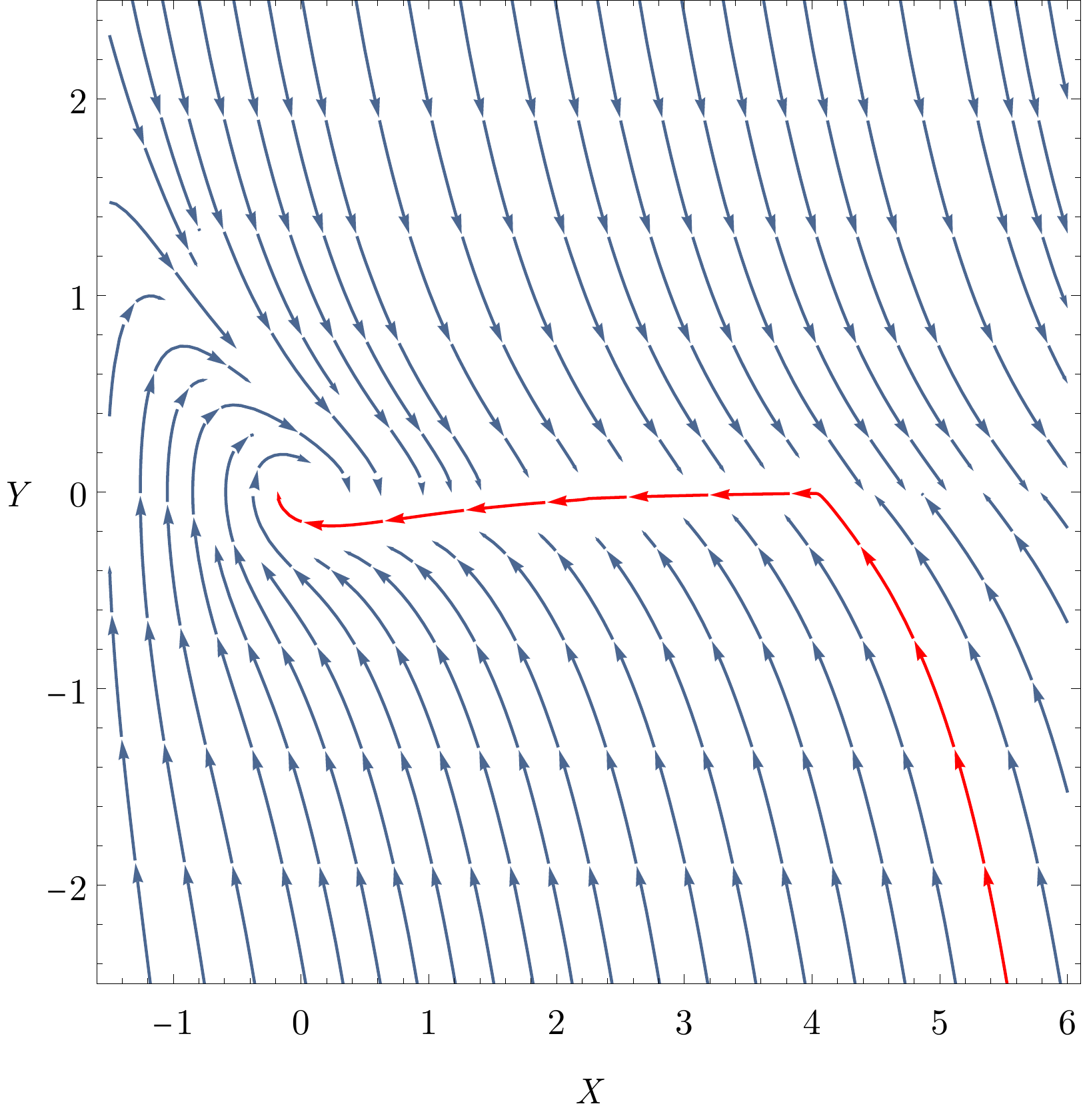}\includegraphics[scale=0.42]{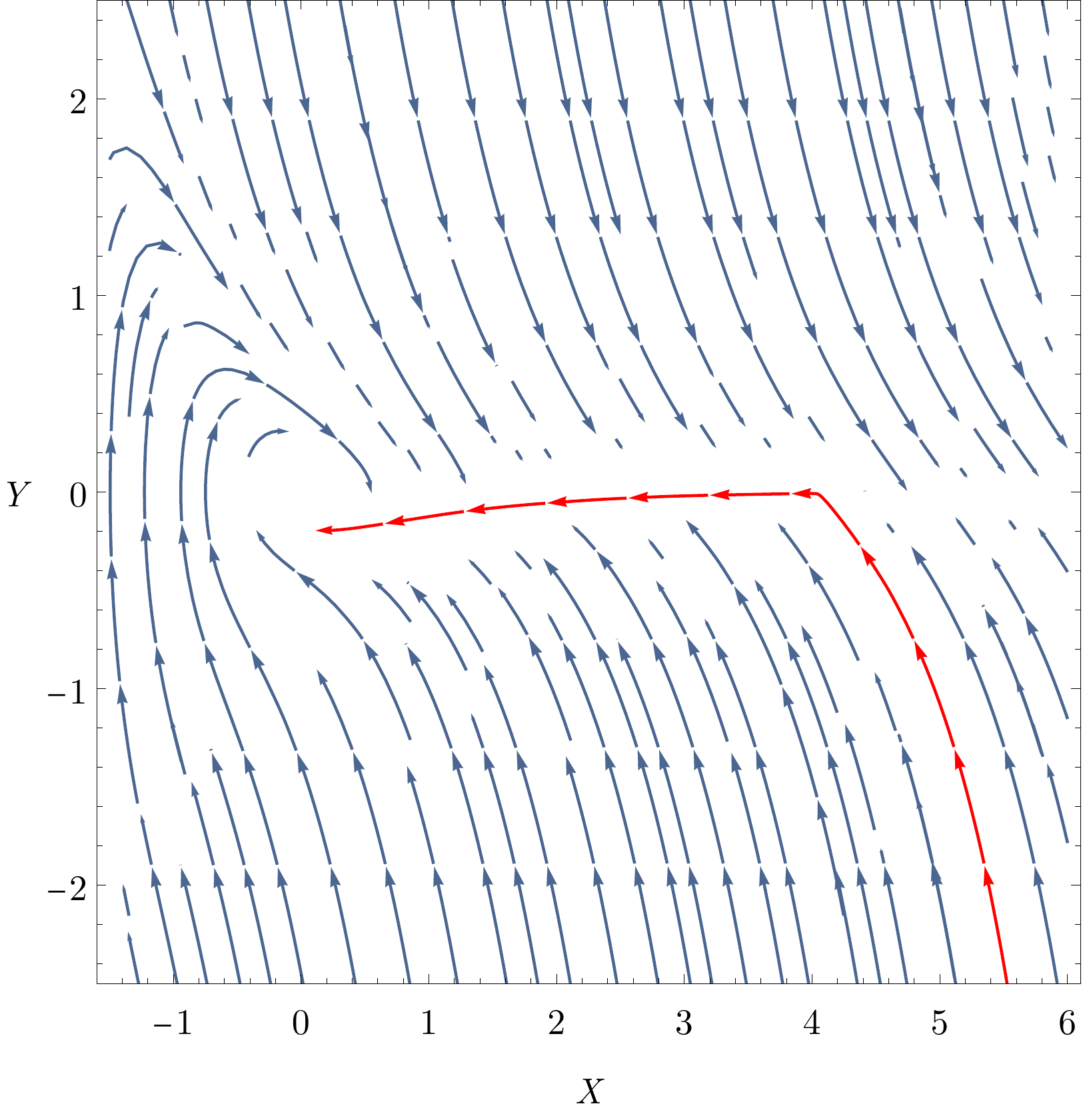}

\caption{Plots of the direction fields on the $\left(X,Y\right)$-phase-space plane
for $\beta_{0}=10^{-3}$ (left) and for $\beta_{0}=10^{-1}$ (right) with
$T=1$ and $S=0.1$. The attractor line solution is present for a
set of values chosen for $T$ and $S$ as large as $-100\lesssim(T,S)\lesssim+100$.
Further details are given in the text.\label{fig:XY}}
\end{figure}

In order to illustrate how the slow-roll regime and the ``graceful
exit\textquotedblright{} take place in our model, we treat $X$ and
$T$ as independent variables and build the direction fields numerically
on the $\left(X,Y\right)$ and $\left(S,T\right)$ planes. With these
assumptions,
\begin{align*}
\frac{\partial Y}{\partial X} & =\frac{j_{1}\left(X,Y,T,S\right)}{Y},\\
\frac{\partial S}{\partial T} & =\frac{j_{2}\left(X,Y,T,S\right)}{\beta_{0}S},
\end{align*}
and we proceed with a numerical analysis summarized in Fig. \ref{fig:XY}.
The direction fields on the $\left(X,Y\right)$-phase-space plane were
built for fixed values of $\beta_{0}$ , $T$ and $S$. The directions
fields on the $\left(T,S\right)$ plane are built for fixed values of
$\beta_{0}$, $X$ and $Y$.


\begin{figure}[h]
\includegraphics[scale=0.3]{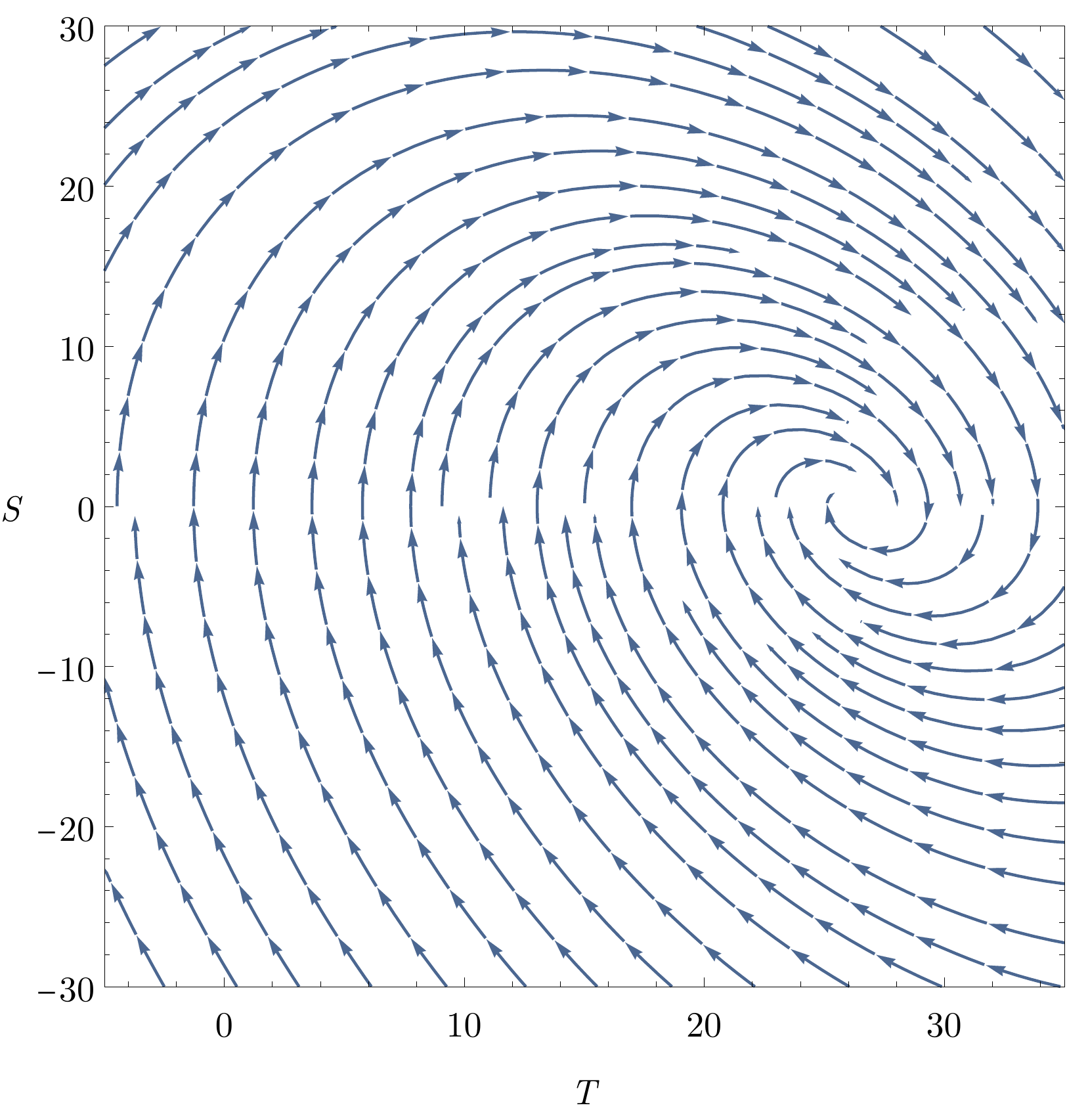}\includegraphics[scale=0.3]{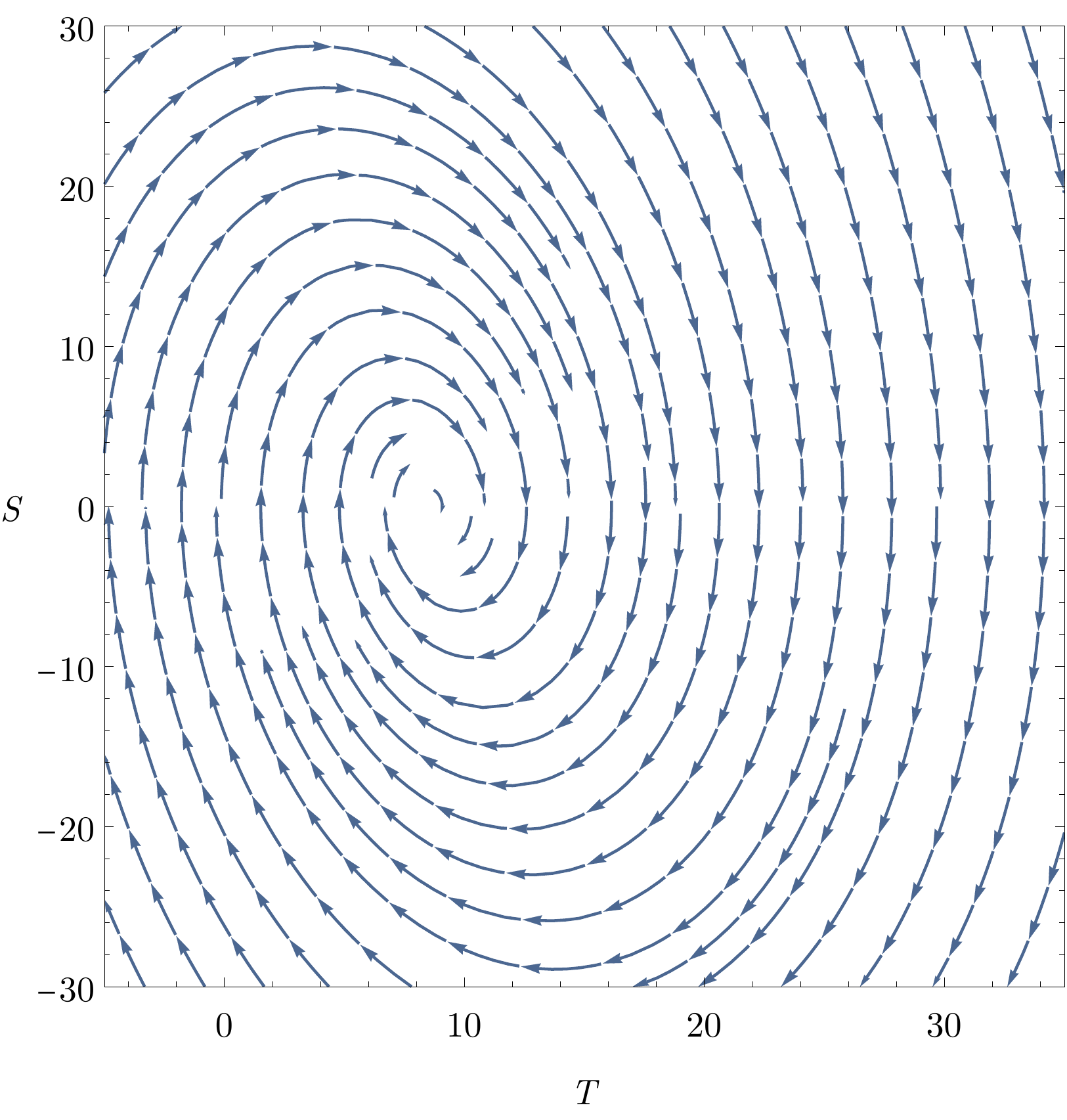}\includegraphics[scale=0.3]{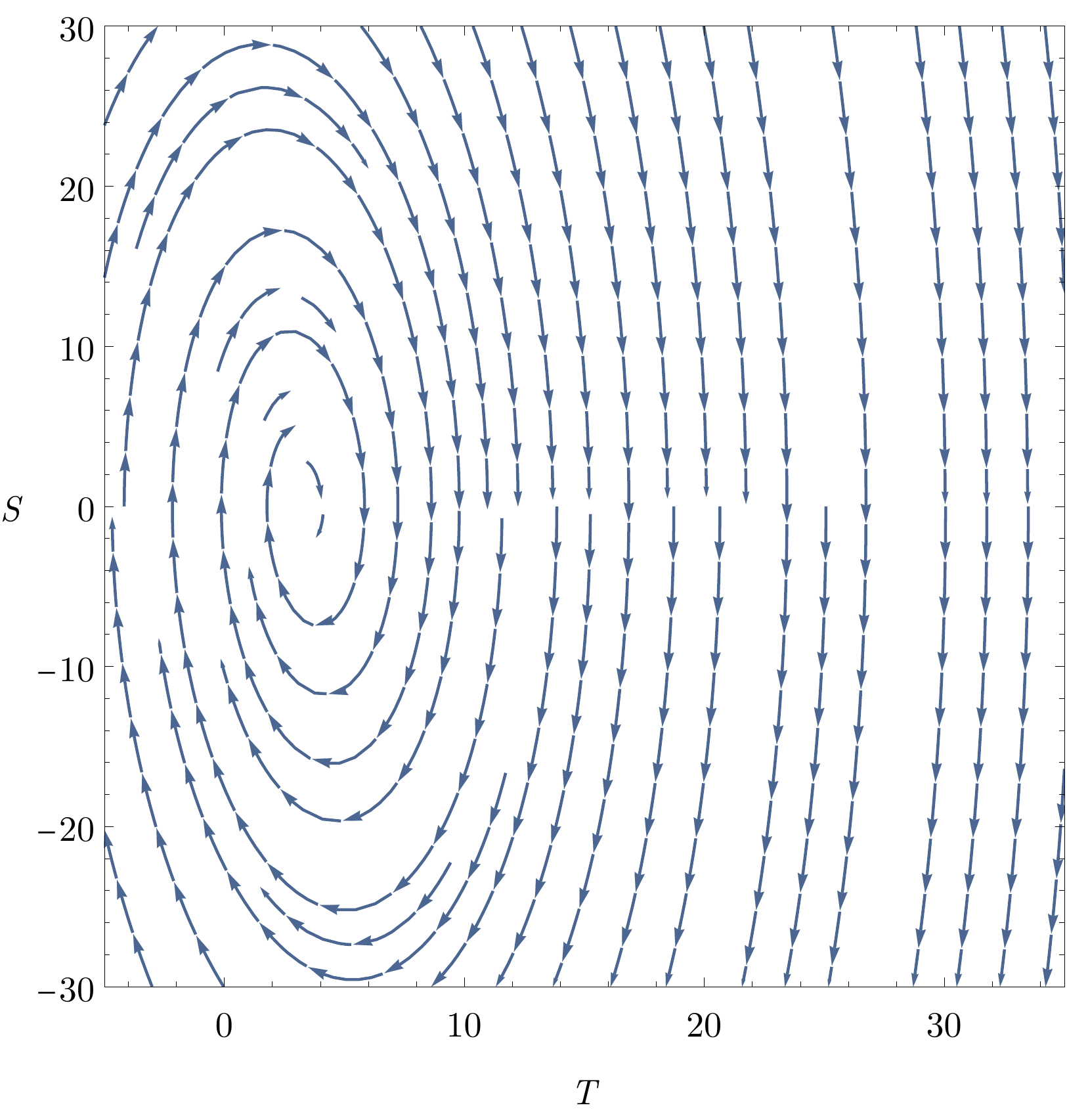}

\caption{Plots of the direction fields on the $\left(T,S\right)$-phase-space plane
for $X=4$ (left)$,$ $X=3$ (middle) and $X=2$ (right) with $\beta_{0}=10^{-2}$.
The small value of $Y$ is set by Eq. (\ref{Phi adm aprox final})\emph{.\label{fig:TS}}}
\end{figure}

The most important feature of the plots shown in Fig. \ref{fig:XY}
is the existence of a horizontal attractor line solution. This solution
is present throughout the range $0\leq\beta_{0}<\frac{3}{4}$ for
 a large range of values of $T$ and $S$, typically $-100\leq\left(T,S\right)\leq100$.
A more complete graphical analysis where we vary $T$ and $S$ also
shows that the horizontal attractor line shifts slightly to the right
for larger values of $\left\vert T\right\vert $ and $\left\vert S\right\vert $.
The direction fields on the $\left(X,Y\right)$-phase space resemble those
obtained in Starobinsky model: We have an attractor region which eventually
leads the trajectories to the origin of the $\left(X,Y\right)$ plane.
The trajectories are slightly different from those of the Starobinky
model (there are small differences on the slope of the attractor solution) but
they are also characterized by positive values of $X$ and by $Y$-coordinates
that are negative and small in magnitude. It is important to realize
that while the trajectories on the $\left(X,Y\right)$ plane evolve in
time in the direction of the point $X=Y=0$, the trajectories on the $\left(S,T\right)$
plane concomitantly evolve to $T=S=0$ (see the sequence of plots
in Fig. \ref{fig:TS}).


In the $\left(T,S\right)$ space (Fig. \ref{fig:TS}), we do not see an
attractor line region as clearly as on the $\left(X,Y\right)$ plane (Fig.
\ref{fig:XY}). However, we verify the existence of an accumulation
point that moves towards the origin as $X$ and $Y$ evolve on the
attractor solution. Besides, this accumulation point \textemdash{}
characterized by a constant value of $S$, namely $S\sim0$ \textemdash{}
is present throughout the range $0\leq\beta_{0}<\frac{3}{4}$.

Therefore, the trajectories on both $\left(X,Y\right)$ and $\left(T,S\right)$
planes allow us to identify an attractor region in a neighbourhood
of which the field equations can be approximated.


\subsubsection{Approximated equations in the attractor region: inflationary regime\label{sec - Approx. equations}}

We start by characterizing the magnitude of $\beta_{0}$ by a parameter
$X^{\ast}$ such that
\[
\beta_{0}\equiv e^{-X^{\ast}}<\frac{3}{4}.
\]
The field equations will be analyzed in the attractor sub-region of
the phase space where $X<X^{\ast}$ which implies $\beta_{0}<e^{-X}$.
\footnote{From Fig. \ref{fig:XY} we see this attractor sub-region always exists.}
In addition, from the discussion around the graphic results (Figs.
\ref{fig:XY} and \ref{fig:TS}) we can assume
\begin{equation}
\left\vert Y\right\vert \ll1\text{ with }Y<0,\label{Y peq}
\end{equation}
and
\begin{equation}
S\ll T,\qquad\left\vert S\right\vert \ll1.\label{S peq}
\end{equation}
If we recall that $\dot{S}=\frac{\partial S}{\partial X}Y$ , $\frac{\partial S}{\partial X}\ll1$,
$\frac{\partial Y}{\partial X}\ll1$ in the attractor region, then
the terms in Eqs. (\ref{Phi adm1}), (\ref{Quisi adm1}) and (\ref{h adm1})
can be compared order by order. As a result, they can be approximated
by
\begin{align}
Y & \simeq-\frac{1}{3}h_{\kappa}^{-1}\frac{e^{-X}}{\left(3-\beta_{0}e^{X}\right)},\label{Phi adm aprox final}\\
T & \simeq-Ye^{2X},\label{Quisi adm aprox final}\\
h_{\kappa}^{2} & \simeq\frac{1}{12}\left(1-2e^{-X}\right).\label{H adm aprox final}
\end{align}
With these equations, variables $Y$ and $T$ can be determined once
$\beta_{0}$ and $X$ are given.

One of the most important results of these approximate solutions lies
on Eq. (\ref{H adm aprox final}): it shows the quasi-exponential
behaviour of the scale factor. This expression also reveals that the
greater the values of $X$ the closer the scale factor behaves
to an exponential growth. Thus, we conclude that the attractor region
corresponds to an inflationary expansion regime.

Several tests were performed to check the consistency of this approximation
with the above numerical results. As a summary, we point out that
the greater the value of $X$ (respecting $X<X^{\ast}$) the better
the above equations will describe the exact results. From a practical
point of view, the above expressions will already constitute an excellent
approximation of the attractor phase for $X\geq2$. For example, we
obtain $T=26.2,$ $8.7$ or $3.4$ for $\beta_{0}=10^{-2}$ and $X=4,3$
or $2$ respectively, showing that the accumulation points in the
plots of Fig. \ref{fig:TS} are very well localized.

Now we are ready to evaluate the slow-roll parameters.

\bigskip{}


\paragraph{Slow roll parameters and number of e-folds:}

In order to accommodate a slow-roll quasi-exponential inflation, any
model must satisfy the following conditions:
\begin{align}
\epsilon_{H} & =-\frac{1}{H^{2}}\frac{dH}{dt}\ll1,\label{epsolon}\\
\eta_{H} & =\frac{1}{H\epsilon}\frac{d}{dt}\left(\frac{1}{H^{2}}\frac{dH}{dt}\right)\ll1.\label{eta}
\end{align}
In our case, the approximations of the field equations around the
slow-roll attractor lead to:
\begin{align}
\epsilon_{H} & \approx\frac{4e^{-2X}}{3\left(1-\frac{\beta_{0}}{3}e^{X}\right)},\label{SR epsolon}\\
\eta_{H} & \approx-\frac{8}{3}\frac{\left(1-\frac{\beta_{0}}{2}e^{X}\right)}{\left(1-\frac{\beta_{0}}{3}e^{X}\right)^{2}}e^{-X}.\label{SR eta}
\end{align}
The denominator of both expressions demand that $\frac{\beta_{0}}{3}e^{X}\neq1$.
Actually, as will be seen below, the approximations demand $\frac{\beta_{0}}{3}e^{X}<1$\emph{
}for consistency with condition $X<X^{\ast}$. Note that the slow-roll
parameters are suppressed by $e^{-X}$ factors. This suggests sorting
all quantities in orders of slow-roll according to the number of factors
$e^{-X}$ they display. Thus, $\epsilon_{H}$ and $\eta_{H}$ are
second- and first-order slow-roll quantities, respectively. This type
of classification will be especially important in the approximation
of perturbative equations.

The number of $e$-folds is now evaluated. As usual \cite{LiddleLyth},
it is defined as
\[
N\equiv\ln\left(\frac{a_{\text{end}}}{a}\right)=\int\limits _{t}^{t_{\text{end}}}Hdt,
\]
where subscript $_{\text{end}}$ corresponds to the end of inflation\emph{.}
The attractor phase imposes a monotonic relation between $X$ and
$t$ during the inflationary regime. Hence, Eq. (\ref{Phi adm aprox final})
can be used to recast $N$ in the form
\[
N\simeq-3\int\limits _{X}^{X_{\text{end}}}h_{\kappa}^{2}\left[\frac{3-\beta_{0}e^{X}}{e^{-X}}\right]dX
\]
The upper limit of this integral is taken as $X_{\text{end}}\simeq0$.
In the slow-roll approximation, the integral gives:
\begin{equation}
N\simeq\frac{3}{4}e^{X}-\frac{1}{8}\beta_{0}e^{2X}.\label{e-folds}
\end{equation}
This equation establishes a relation between $e^{X}$ and the number
of e-folds, which can be used to write the former as function of the
latter. Since this is a second order equation for $e^{X}$, two solutions
are found:
\[
e^{X}=\frac{3}{\beta_{0}}\left[1\pm\sqrt{1-\frac{8}{9}\beta_{0}N}\right]
\]
The ``$+$\textquotedblright{} sign must be discarded, should our
model restore Starobinsky's results in the limit $\beta_{0}\rightarrow0$.
That is what will be assumed henceforth:
\begin{equation}
e^{X}=\frac{3}{\beta_{0}}\left[1-\sqrt{1-\frac{8}{9}\beta_{0}N}\right].\label{e-folds e X}
\end{equation}
It is clear from this expression that real values for $e^{X}$ are
obtained only if $N\leq\frac{9}{8\beta_{0}}$. This fixes an upper
limit for $N$ and, consequently, for $e^{X}$:
\[
N_{\max}=\frac{9}{8\beta_{0}}\Rightarrow e^{X_{\max}}=\frac{3}{\beta_{0}}.
\]
These values cannot be physically attained and should be considered
solely as constraints, since they actually provoke the divergence
of the slow-roll parameters violating the conditions for inflation.
From these results, it is clear that the maximum number of e-folds and
$X_{\max}$ are determined given a value for $\beta_{0}$.
We will use the above results in the following way: Given physical
limits for $N$, we expect to set physical limits to $\beta_{0}$.
As we can see from Eq. (\ref{e-folds e X}),
\[
e^{X}<\frac{3}{\beta_{0}}\Rightarrow\beta_{0}<\frac{9}{8N}.
\]
Observationally, it is usually expected $N\geq50$. Thus, we must
have $\beta_{0}<0.0225$ for consistency.


\section{Perturbed cosmological equations\label{sec:Perturbations}}

An important feature of the inflationary paradigm is to engender the
primordial seeds responsible for the large-scale structures formation
observed in our universe. Usually, these seeds are generated from
small quantum fluctuations in a homogeneous and isotropic background
during the inflationary regime. Thus, in order to study the characteristics
of these fluctuations in the context of our model, it is necessary
to perturb the cosmological field equations obtained in Section \ref{sec - MGA}.

The fundamental quantities to be perturbed are:
\begin{align*}
\Phi & =\Phi_{\left(0\right)}+\delta\Phi,\\
\xi^{\mu} & =\xi_{\left(0\right)}^{\mu}+\delta\xi^{\mu},\\
g_{\mu\nu} & =g_{\mu\nu\left(0\right)}+\delta g_{\mu\nu},
\end{align*}
where index $\left(0\right)$ indicates a background quantity. Vector
and tensor perturbations ($\delta\xi^{\mu}$ and $\delta g_{\mu\nu}$) can
be decomposed into irreducible scalar-vector-tensor perturbations
(SVT decomposition). Thus, using the notation defined in Eq. (\ref{Var adm}),
it is possible to write the above quantities as
\begin{align}
\delta\Phi & =M_{Pl}\sqrt{\frac{3}{2}}\delta X,\label{Def Phi em X}\\
\delta\xi^{0} & =\kappa_{0}^{3/2}\delta T,\label{Def xi_0 em T}\\
\delta\xi^{i} & =\kappa_{0}^{3/2}\left(\frac{1}{\kappa_{0}^{1/2}}\left(\partial^{i}\delta W\right)+\delta V^{i}\right),\label{Def xi_j em W e Vj}
\end{align}
and
\begin{align}
\delta g_{00} & =-2\delta A,\label{Def pert metric 00}\\
\delta g_{0i} & =\frac{1}{\kappa_{0}^{1/2}}\left(\partial_{i}\delta B\right)+\delta B_{i},\label{Def pert metric 0i}\\
\delta g_{ij} & =a^{2}\left[2\delta_{ij}\delta C+\frac{2}{\kappa_{0}}\left(\partial_{i}\partial_{j}\delta E\right)+\frac{1}{\kappa_{0}^{1/2}}\partial_{i}\delta E_{j}+\frac{1}{\kappa_{0}^{1/2}}\partial_{j}\delta E_{i}+2\delta E_{ij}\right],\label{Def pert metric ij}
\end{align}
where $\partial^{i}=g_{\left(0\right)}^{ij}\partial_{j}$. The $\kappa_{0}^{1/2}$
factors were included to make all perturbations dimensionless. Notice
that the perturbation $\delta\xi^{\mu}$ is decomposed via SVT in
two scalar degrees of freedom (namely, $\ensuremath{\delta T}$ and $\ensuremath{\delta W}$) and a vectorial one ($\delta V^{i}$). This decomposition is analogous
to the one performed for $\delta g_{0\mu}$, cf. Eqs.~(\ref{Def pert metric 00})
and (\ref{Def pert metric 0i}).

The complete line element reads:
\begin{align}
ds^{2} & =-\left(1+2\delta A\right)dt^{2}+\left[\frac{2}{\kappa_{0}^{1/2}}\left(\partial_{i}\delta B\right)+2\delta B_{i}\right]dtdx^{i}\nonumber \\
 & +a^{2}\left[\left(1+2\delta C\right)\delta_{ij}+\frac{2}{\kappa_{0}}\partial_{i}\partial_{j}\delta E+\frac{1}{\kappa_{0}^{1/2}}\left(\partial_{i}\delta E_{j}+\partial_{j}\delta E_{i}\right)+2\delta E_{ij}\right]dx^{i}dx^{j}.\label{metrica perturbada}
\end{align}
Consequently, there are seven scalar perturbed quantities ($\delta X$,
$\delta T$, $\delta W$, $\delta A$, $\delta B$, $\delta C$ and
$\delta E$), three divergenceless vector perturbations ($\delta V^{i}$,
$\delta B_{i}$ and $\delta E_{i}$) and one transverse-traceless
tensor perturbation ($\delta E_{ij}$).

In addition to these eleven fundamental perturbed quantities, it is
also adequate to introduce auxiliary perturbations associated with
the energy-momentum tensor of an imperfect fluid \textemdash{} Eq.
(\ref{Tmunu imperfect}). In effect, we shall consider the four perturbed
quantities $\delta\varepsilon$, $\delta p$, $\delta u^{\mu}$ and
$\delta q_{\mu}$ coming from Eq. (\ref{Tmunu imperfect}) with null
viscous shear tensor. Under the constraints $u^{\mu}u_{\mu}=-1$ and
$u^{\mu}q_{\mu}=0$, perturbations $\delta u^{\mu}$ and $\delta q_{\mu}$
can be decomposed as:
\begin{align}
\delta u^{0} & =-\delta A,\label{4 - v0}\\
\delta u^{i} & =\frac{1}{\kappa_{0}^{1/2}}\partial^{i}\delta v+\delta w^{i},\label{4 - vi}\\
\delta q_{0} & =0,\label{4 - q0}\\
\delta q_{i} & =\frac{1}{\kappa_{0}^{1/2}}\partial_{i}\delta q+\delta r_{i},\label{4 - qi}
\end{align}
where $\delta w^{i}$ and $\delta r_{i}$ are vectors of zero divergence.
It is noteworthy that scalar, vector and tensor perturbations evolve
independently in the linear regime; therefore each set can be treated
separately.


\subsection{Scalar equations}

In the linear regime of perturbations there are six scalar field equations:
One associated with the scalaron $\Phi$, two related to the vector
field $\xi^{\mu}$ and three coming from Einstein equations.

By perturbing Eq. (\ref{Eq Phi sem def})\emph{ }we obtain, after
an extensive manipulation, the expression
\begin{align}
2\dot{Y}\delta A+Y\left(\delta\dot{A}-3\delta\dot{C}-\bar{\partial}_{0}\left(a^{2}\bar{\nabla}^{2}\delta E\right)\right)+6h_{\kappa}Y\delta A+\frac{Y}{\sqrt{3}}\bar{\nabla}^{2}\left(\delta B\right)\nonumber \\
-3h_{\kappa}\delta\dot{X}-\delta\ddot{X}+\bar{\nabla}^{2}\left(\delta X\right)+\frac{1}{3}e^{-X}\delta X\left(1-2e^{-X}\right)\nonumber \\
+\frac{\beta_{0}}{3}e^{-2X}f\left(\delta X,\delta T,\delta W,\delta A,\delta C,\delta E\right)=0,\label{Eq completa Phi var adm}
\end{align}
where
\[
f\left(\delta X,\delta T,\delta W,\delta A,\delta C,\delta E\right)=
\]
\begin{align}
= & \left[e^{X}+2\beta_{0}\left(3h_{\kappa}T-2TY+S\right)\right]\left[\left(3h_{\kappa}T-2TY+S\right)\delta X+\left(2Y-3h_{\kappa}\right)\delta T+2T\delta\dot{X}\right]\nonumber \\
 & -\left[e^{X}+2\beta_{0}\left(3h_{\kappa}T-2TY+S\right)\right]\left[\delta\dot{T}+\bar{\nabla}^{2}\left(\delta W\right)+T\left(\delta\dot{A}+3\delta\dot{C}+\bar{\partial}_{0}\left(a^{2}\bar{\nabla}^{2}\delta E\right)\right)\right]\nonumber \\
 & -Te^{-X}\left(\frac{3}{2}T\delta X-\delta T-T\delta A\right).\label{f def}
\end{align}
The dimensionless barred operator is defined as:
\[
\bar{\nabla}^{2}Q\equiv\frac{1}{\kappa_{0}a^{2}}\delta^{ji}\partial_{j}\partial_{i}Q.
\]
Notice that only the third line in Eq. (\ref{Eq completa Phi var adm})
corresponds to corrections to Starobinsky inflation.

The perturbed equations associated with $\xi^{0}$ and $\xi^{i}$,
Eq. (\ref{Eq Psi sem def 1}), lead to
\begin{gather}
\bar{\partial}_{0}\left[e^{X}\delta X+\beta_{0}\left[\left(3h_{\kappa}-2Y\right)\delta T+\delta\dot{T}-2T\delta\dot{X}+\bar{\nabla}^{2}\left(\delta W\right)+T\left(\delta\dot{A}+3\delta\dot{C}+\bar{\partial}_{0}\left(a^{2}\bar{\nabla}^{2}\delta E\right)\right)\right]\right]\nonumber \\
-e^{-X}\left(T\delta X-2T\delta A-\delta T\right)=0\label{Eq escalar Csi 0}
\end{gather}
and
\begin{gather}
\partial_{j}\left[e^{X}\delta X+\beta_{0}\left[\left(3h_{\kappa}-2Y\right)\delta T+\delta\dot{T}-2T\delta\dot{X}+\bar{\nabla}^{2}\left(\delta W\right)+T\left(\delta\dot{A}+3\delta\dot{C}+\bar{\partial}_{0}\left(a^{2}\bar{\nabla}^{2}\delta E\right)\right)\right]\right]\nonumber \\
-e^{-X}\partial_{j}\left[\left(T\delta B+\delta W\right)\right]=0.\label{Eq escalar Csi j}
\end{gather}
By combining these two equations we can obtain a simpler expression,
given by
\begin{equation}
\partial_{j}\left[T\delta X-\delta T+Y\delta W-\delta\dot{W}-2T\delta A+YT\delta B-S\delta B-T\delta\dot{B}\right]=0.\label{Eq combinada}
\end{equation}

It is also necessary to perturb Einstein equations. These equations,
in a gauge invariant form \cite{Mukhanov}, are given by
\begin{align}
\bar{\nabla}^{2}\Psi-3h_{\kappa}\left(\dot{\Psi}+h_{\kappa}\Psi\right) & =-\frac{\widetilde{\delta\varepsilon}}{2\kappa_{0}M_{Pl}^{2}},\label{Eins Inv gauge esc 1}\\
\dot{\Psi}+h_{\kappa}\Psi & =\frac{1}{2\kappa_{0}M_{Pl}^{2}}\left[\left(\varepsilon+p\right)\widetilde{\delta v}+\widetilde{\delta q}\right],\label{Eins Inv gauge esc 2}\\
\ddot{\Psi}+4h_{\kappa}\dot{\Psi}+\left(2\dot{h}_{\kappa}+3h_{\kappa}^{2}\right)\Psi & =-\frac{\widetilde{\delta p}}{2\kappa_{0}M_{Pl}^{2}},\label{Eins Inv gauge esc 3}
\end{align}
where the choice of different gauges can be made through the expressions:
\begin{align}
\Psi & =-\delta A-\bar{\partial}_{0}\left(\delta B-a^{2}\delta\dot{E}\right),\label{gauge Phi}\\
\widetilde{\delta\varepsilon} & =\delta\varepsilon+\dot{\varepsilon}_{\left(0\right)}\left(\delta B-a^{2}\delta\dot{E}\right),\label{gauge dens de energia}\\
\widetilde{\delta p} & =\delta p+\dot{p}_{\left(0\right)}\left(\delta B-a^{2}\delta\dot{E}\right),\label{gauge pressao}\\
\widetilde{\delta v} & =\delta v+a^{2}\delta\dot{E},\label{gauge velocidade}\\
\widetilde{\delta q} & =\delta q.\label{gauge fluxo de calor}
\end{align}

The last equation states that the heat flux $\delta q$ is naturally gauge
invariant. In addition to Eqs. (\ref{Eins Inv gauge esc 1}), (\ref{Eins Inv gauge esc 2})
and (\ref{Eins Inv gauge esc 3}), we have the constraint
\begin{equation}
\delta A+\delta C=-\frac{1}{a}\bar{\partial}_{0}\left[a\left(\delta B-a^{2}\delta\dot{E}\right)\right],\label{vinculo}
\end{equation}
arising from Einstein's equation $\delta G_{ij}=M_{Pl}^{-2}\delta T_{ij}$
with $i\neq j$. The relationship between the quantities $\delta\varepsilon$,
$\delta p$, $\delta v$ and $\delta q$ and the fundamental scalar
perturbations are obtained from the Eqs. (\ref{epsolon mais p FI}),
(\ref{4 velocidade FI}) and (\ref{fluxo de calor FI}).\emph{ }By
perturbing these equations we obtain
\begin{align}
\delta\varepsilon+\delta p & =3\kappa_{0}M_{Pl}^{2}\left[Y\delta\dot{X}-Y^{2}\delta A+\frac{\beta_{0}}{3}e^{-3X}T\left(\frac{3}{2}T\delta X-\delta T-T\delta A\right)\right],\label{Pert de epsilon mais p}\\
\delta v & =-\frac{\delta X+Y\delta B+\sqrt{\frac{\beta_{0}e^{-3X}}{3}}\delta W}{\left(Y-\sqrt{\frac{\beta_{0}e^{-3X}}{3}}T\right)},\label{Pert v}\\
\delta q & =\frac{3}{2}\sqrt{\frac{\beta_{0}e^{-3X}}{3}}\kappa_{0}M_{Pl}^{2}\left(T\delta X+YT\delta B+Y\delta W\right).\label{Pert q}
\end{align}
These last three equations together with the perturbation for $\delta\varepsilon$
\textemdash{} see Appendix \ref{sec - pert of energy density} \textemdash{}
complete the description of the perturbed Einstein's equations.

The set of equations (\ref{Eq completa Phi var adm},
\ref{Eq escalar Csi 0}, \ref{Eq escalar Csi j}, \ref{Eins Inv gauge esc 1},
\ref{Eins Inv gauge esc 2}, \ref{Eins Inv gauge esc 3}, \ref{vinculo})
establishes the dynamics of the scalar perturbations.
Let us emphasize that not all of these perturbations are dynamical quantities.
Actually, a quick analyzis of the Cauchy problem shows that only four
of these equations are truly dynamical equations, while three of them
constitute constraints between the variables. It is interesting to note
that the number of degrees of freedom in our higher-order scalar-vector approach is in agreement with the number of degrees of freedom in the higher-order two-scalar approach of Ref. \cite{Castellanos2018}. In the later case, besides the perturbations of the two scalar fields, there are two scalar perturbations from the metric.

Here, two of the seven scalar perturbations can
be ``eliminated\textquotedbl{} by an appropriate gauge choice. Moreover,
Eq. (\ref{vinculo}) allows us to write either $\delta C$ or $\delta A$
in terms of the other three metric perturbations. Thus, the problem
is completely characterized by four differential equations. Since
the expressions for $\delta\varepsilon$ and $\delta p$ contain a
lot of terms, it is convenient to select a set of equations avoiding
these perturbations. A natural choice here is to work with Eqs. (\ref{Eq completa Phi var adm}),
(\ref{Eq escalar Csi j}), (\ref{Eq combinada}) and (\ref{Eins Inv gauge esc 2}).
This will be done in Section \ref{sec - sol. scalar fields} with
the use of slow-roll approximation.


\subsection{Vector and tensor equations}

There are three fundamental equations associated with vector perturbations.
The first one is obtained by perturbing Eq. (\ref{Eq Psi sem def 1});
the result is:
\begin{equation}
\delta V_{i}=-T\delta B_{i}.\label{vinculo vetorial}
\end{equation}
The other two come from perturbations in $0i$- and $ij$-components
of Einstein's equations:
\begin{align}
\bar{\nabla}^{2}\delta F_{i} & =\frac{2}{a\kappa_{0}M_{Pl}^{2}}\left[\left(\varepsilon+p\right)\widetilde{\delta w_{i}}+\widetilde{\delta r_{i}}\right],\label{Eins Inv gauge vec 1}\\
\partial_{i}\delta\dot{F}_{j}+\partial_{j}\delta\dot{F}_{i}+h_{\kappa}\left(\partial_{i}\delta F_{j}+\partial_{j}\delta F_{i}\right) & =0,\label{Eins Inv gauge vec 2}
\end{align}
where
\begin{align}
\delta F_{i} & =\delta B_{i}-a^{2}\delta\dot{E}_{i},\label{gauge Fi}\\
\widetilde{\delta w_{i}} & =\delta w_{i}+\delta B_{i},\label{gauge wi}\\
\widetilde{\delta r_{i}} & =\delta r_{i},\label{gauge ri}
\end{align}
are gauge invariant quantities. Due to the constraint (\ref{vinculo vetorial}),
it is possible to show that
\begin{equation}
\widetilde{\delta w_{i}}=\widetilde{\delta r_{i}}=0,\label{all zero}
\end{equation}
i.e. the effective energy-momentum tensor (\ref{Tmunu sem definicao})
has no vector perturbations. This result was expected since the term
$\nabla_{\mu}R\nabla^{\mu}R$ responsible for the extra vector perturbations
can be written as $-R\square R$, which contains only scalar degrees
of freedom (see \cite{Wands1994}).

Finally, there is only one equation associated with the tensor degree
of freedom:
\begin{equation}
\delta\ddot{E}_{ij}+3h_{\kappa}\delta\dot{E}_{ij}-\bar{\nabla}^{2}\delta E_{ij}=0.\label{Eins Inv gauge tensor}
\end{equation}
This equation is derived from $\delta G_{ij}=M_{Pl}^{-2}\delta T_{ij}$
and represents gravitational waves freely propagating in a homogeneous
and isotropic background.


\section{Solutions of the perturbed cosmological equations \label{sec:Sol_pert_eq}}

The equations derived in the previous section are complicated. However,
in the attractor region, where the slow-roll approximation is valid,
these equations are considerably simplified and they can be treated
analytically.


\subsection{Scalar solutions\label{sec - sol. scalar fields}}

The implementation of approximations in the scalar equations should
take into account that, in general, different perturbations in a given
gauge have different orders of slow-roll. For example, in the Newtonian
gauge, Eq. (\ref{Eins Inv gauge esc 2}) for Starobinsky inflation
($\beta_{0}=0$) is written as
\[
\delta\dot{A}+h_{\kappa}\delta A=\frac{3}{4}Y\delta X.
\]
During the inflationary regime, where $h_{\kappa}\sim1$, this equation
tells us $\delta A\sim Y\delta X\sim e^{-X}\delta X$, which means
that the metric perturbation is a slow-roll factor smaller than the
scalar field perturbation.

For the case $\beta_{0}\neq0$, the situation is more complicated
because Eqs. (\ref{Eq escalar Csi j}) and (\ref{Eq combinada}) indicate
that the perturbations $\delta T$ and $\delta W$ are different from
$\delta X$ concerning the order of slow-roll factors. This can
be explicitly seen by writing
Eq. (\ref{Eq escalar Csi j}) in the Newtonian gauge\footnote{The derivative $\partial_{j}$ disappears because this equation must
be satisfied independently for each $k^{j}$ mode.}
\begin{equation}
e^{X}\delta X+\beta_{0}\left[\left(3h_{\kappa}-2Y\right)\delta T+\delta\dot{T}-2T\delta\dot{X}+\bar{\nabla}^{2}\left(\delta W\right)-2T\delta\dot{A}\right]-e^{-X}\left(\delta W\right)=0.\label{Eq escalar Csi j 1}
\end{equation}
As in the attractor sub-region $\beta_{0}\lesssim e^{-X}$ and $T\sim e^{X}$
(see Section \ref{sec - Approx. equations}), this equation tells
us that $\delta T$ or $\delta W$ must be two slow-roll factors larger
than $\delta X$. Moreover, Eq. (\ref{Eq combinada})
\begin{equation}
T\delta X-\delta T+Y\delta W-\delta\dot{W}-2T\delta A=0\label{Eq combinada 1}
\end{equation}
shows that $\delta T$ and $\delta W$ are of the same order in slow-roll.
Thus, in the Newtonian gauge, Eqs.~(\ref{Eq escalar Csi j 1}), (\ref{Eq combinada 1})
and (\ref{Eins Inv gauge esc 2}) suggest that
\begin{align}
\delta T & \sim\delta W\sim e^{2X}\delta X,\label{consis 1}\\
\delta A & \sim e^{-X}\delta X.\label{consis 2}
\end{align}

The next step is to use (\ref{consis 1}) and (\ref{consis 2}) to
simplify the expressions (\ref{Eq completa Phi var adm}), (\ref{Eq escalar Csi j})
and (\ref{Eq combinada}). During the inflationary regime:
\[
T\sim e^{X}\text{, }Y\sim e^{-X}\text{, }h_{\kappa}\sim1\text{\ e }\beta_{0}\lesssim e^{-X}\text{\ .}
\]
So, up to slow-roll leading order, Eqs.~(\ref{Eq completa Phi var adm}),
(\ref{Eq escalar Csi j}) and (\ref{Eq combinada}) are approximated
by:
\begin{align}
\delta\ddot{X}+3h_{\kappa}\delta\dot{X}-\bar{\nabla}^{2}\left(\delta X\right)+\frac{\beta_{0}}{3}e^{-X}\left[3h_{\kappa}\delta T+\delta\dot{T}+\bar{\nabla}^{2}\left(\delta W\right)\right] & \simeq0,\label{Aux1}\\
\delta X-e^{-2X}\delta W+\beta_{0}e^{-X}\left[3h_{\kappa}\delta T+\delta\dot{T}+\bar{\nabla}^{2}\left(\delta W\right)\right] & \simeq0,\label{Aux2}\\
\delta\dot{W}+\delta T & \simeq0.\label{Aux3}
\end{align}
In the Starobinsky limit Eqs.~(\ref{Aux2}) and (\ref{Aux3}) are
not present and Eq.~(\ref{Aux1}) reduces to the usual expression
for a single scalar field. The combination of the three previous equations
results in
\begin{align}
\delta\ddot{X}+3h_{\kappa}\delta\dot{X}-\bar{\nabla}^{2}\left(\delta X\right) & \simeq\frac{1}{3}\left(\delta X-e^{-2X}\delta W\right),\label{delta X aprox}\\
\beta_{0}e^{-X}\left[\delta\ddot{W}+3h_{\kappa}\delta\dot{W}-\bar{\nabla}^{2}\left(\delta W\right)\right] & \simeq\delta X-e^{-2X}\delta W.\label{delta W aprox}
\end{align}
In the slow-roll leading-order approximation, background terms can
be considered constant with respect to time derivatives, i.e.
\[
\bar{\partial}_{0}\left(e^{-X}\delta W\right)=\delta W\bar{\partial}_{0}\left(e^{-X}\right)+e^{-X}\bar{\partial}_{0}\left(\delta W\right)\simeq e^{-X}\bar{\partial}_{0}\left(\delta W\right).
\]

Let
\[
\delta\varphi_{1}\equiv a\delta X\text{ e }\delta\varphi_{2}\equiv ae^{-2X}\delta W
\]
Eqs.~(\ref{delta X aprox}) and (\ref{delta W aprox}) then turn to
\begin{align}
\delta\varphi_{1}^{\prime\prime}+\left(k^{2}-\frac{a^{\prime\prime}}{a}\right)\delta\varphi_{1} & \simeq\frac{\kappa_{0}a^{2}}{3}\left(\delta\varphi_{1}-\delta\varphi_{2}\right),\label{delta phi 1 aux}\\
\beta_{0}e^{X}\left[\delta\varphi_{2}^{\prime\prime}+\left(k^{2}-\frac{a^{\prime\prime}}{a}\right)\delta\varphi_{2}\right] & \simeq\kappa_{0}a^{2}\left(\delta\varphi_{1}-\delta\varphi_{2}\right),\label{delta phi 2 aux}
\end{align}
where prime ($'$) indicates derivative with respect to the conformal
time $\eta$. Notice that by introducing the factor $e^{-2X}$ in
the definition of $\delta\varphi_{2}$ we assure $\delta\varphi_{1}$
and $\delta\varphi_{2}$ are of the same slow-roll order. Moreover,
a (quasi-)de Sitter spacetime satisfies $12H^{2}\simeq\kappa_{0}$;
then,
\[
a\simeq-\frac{1}{H\eta}\Rightarrow\frac{a^{\prime\prime}}{a}\simeq2a^{2}H^{2}\simeq\frac{2}{\eta^{2}}.
\]
and Eqs.~(\ref{delta phi 1 aux}) and (\ref{delta phi 2 aux}) lead
to
\begin{align}
\delta\varphi_{1}^{\prime\prime}+k^{2}\left(1-\frac{2}{\eta^{2}k^{2}}\right)\delta\varphi_{1} & \simeq\frac{4}{k^{2}\eta^{2}}k^{2}\left(\delta\varphi_{1}-\delta\varphi_{2}\right),\label{delta phi 1 conforme}\\
\beta_{0}e^{X}\left[\delta\varphi_{2}^{\prime\prime}+k^{2}\left(1-\frac{2}{\eta^{2}k^{2}}\right)\delta\varphi_{2}\right] & \simeq\frac{12}{k^{2}\eta^{2}}k^{2}\left(\delta\varphi_{1}-\delta\varphi_{2}\right).\label{delta phi 2 conforme}
\end{align}
The solution to the above pair of equations depends on initial conditions
deep in the sub-horizon regime, i.e. for $k\eta\gg1$. In this case,
\[
\delta\varphi_{1,2}^{\prime\prime}+k^{2}\delta\varphi_{1,2}\simeq0, \qquad k\eta\gg1,
\]
and the initial conditions $\left(\delta\varphi_{1i},\delta\varphi_{2i}\right)$
are the same because they come from the quantization of identical
equations. Condition $\delta\varphi_{1i}=\delta\varphi_{2i}$ causes
the vanishing of the right-hand side of Eqs. (\ref{delta phi 1 conforme})
and (\ref{delta phi 2 conforme}) for all $\eta$. For this reason,
the evolution of $\left(\delta\varphi_{1},\delta\varphi_{2}\right)$
is dictated by Mukhanov-Sasaki equation
\[
\delta\varphi_{1,2}^{\prime\prime}+k^{2}\left(1-\frac{2}{\eta^{2}k^{2}}\right)\delta\varphi_{1,2}\simeq0.
\]
Therefore, we get a tracking solution:\emph{ }
\begin{equation}
\delta\varphi_{1}=\delta\varphi_{2}\Rightarrow\delta X\simeq e^{-2X}\delta W.\label{vinculo delta X e delta W}
\end{equation}

Few remarks are in order. First, the tracking solution corresponds
to an adiabatic solution. In fact, the heat flux $\delta q$ vanishes
in slow-roll leading order once Eq. (\ref{Pert q}) in Newtonian gauge
reads:
\[
\delta q\simeq\frac{3M_{Pl}^{2}\kappa_{0}}{2}\sqrt{\frac{\beta_{0}e^{X}}{3}}Y\left(e^{-2X}\delta W-\delta X\right)\simeq0.
\]
This is the reason for the comoving curvature being conserved at super-horizon
scales (cf. Appendix \ref{sec - Conservacao R}).

A second remark is: The tracking solution follows the attractor trajectory
defined by the background fields in the phase space. This is checked
by taking the time derivative of (\ref{vinculo delta X e delta W})
\[
\delta\dot{W}\simeq e^{2X}\delta\dot{X}
\]
and using Eqs. (\ref{Var adm}) and (\ref{Aux3}). Then,
\[
\delta T\simeq-e^{2X}\delta Y,
\]
which is of the same type as (\ref{Quisi adm aprox final}). The fact
that the tracking solution follows the attractor line is not so surprising
because the multi-field adiabatic perturbations in inflationary models
are defined as the ones remaining along the background trajectory
of the homogeneous and isotropic fields \cite{Wand2008}.

Finally, something should be said about what happens to the solutions
of Eqs.~(\ref{delta phi 1 conforme}) and (\ref{delta phi 2 conforme})
in the case where the initial conditions $\delta\varphi_{1i}$ and
$\delta\varphi_{2i}$ are different.\footnote{Higher-order slow-roll terms can introduce non-adiabatic initial perturbations.}
In order to perform this analysis, it is convenient to cast (\ref{delta phi 1 conforme})
and (\ref{delta phi 2 conforme}) in terms of the reset scale factor
such that a given scale crosses the horizon $a=1$, i.e. $k=H$. In
this case,
\begin{align}
\delta\varphi_{1}^{\ast\ast}+\frac{2}{a}\delta\varphi_{1}^{\ast}+\left(\frac{1}{a^{4}}-\frac{2}{a^{2}}\right)\delta\varphi_{1} & \simeq\frac{4}{a^{2}}\left(\delta\varphi_{1}-\delta\varphi_{2}\right),\label{eq 1}\\
\beta_{0}e^{X}\left[\delta\varphi_{2}^{\ast\ast}+\frac{2}{a}\delta\varphi_{2}^{\ast}+\left(\frac{1}{a^{4}}-\frac{2}{a^{2}}\right)\delta\varphi_{2}\right] & \simeq\frac{12}{a^{2}}\left(\delta\varphi_{1}-\delta\varphi_{2}\right),\label{eq 2}
\end{align}
where {*} denotes differentiation with respect to the scale factor
and $\beta_{0}e^{X}$ is given by (\ref{e-folds e X}). Eqs.~(\ref{eq 1})
and (\ref{eq 2}) can be studied for different sets of initial conditions
$\left\{ \delta\varphi_{1i}\text{;}\delta\varphi_{2i}\right\} $ and
$\left\{ \delta^{\ast}\varphi_{1i}\text{;}\delta^{\ast}\varphi_{2i}\right\} $
assuming that $k$ crosses the horizon in the interval $50\leq N\leq60$.
A numerical procedure showed the differences between $\delta\varphi_{1}$
and $\delta\varphi_{2}$ are never amplified; furthermore, they are
suppressed by the expansion in the super-horizon regime ($a>1$).
We conclude that any eventual non-adiabatic perturbation generated
by higher-order corrections may be neglected in slow-roll leading
order.

In view of the considerations above, we state that perturbations $\delta X$
and $\delta W$ have the same dynamics \emph{in first order in slow-roll},
both being described by Mukhanov-Sasaki-type equations
\begin{align}
\delta\ddot{X}+3h_{\kappa}\delta\dot{X}-\bar{\nabla}^{2}\left(\delta X\right) & \simeq0,\label{Mu-Sa X}\\
\delta\ddot{W}+3h_{\kappa}\delta\dot{W}-\bar{\nabla}^{2}\left(\delta W\right) & \simeq0.\label{Mu-Sa W}
\end{align}
It is now necessary to decide on which variable is to be quantized.
This variable is associated to the comoving curvature perturbation
\begin{equation}
\mathcal{R}\equiv\delta C+h_{\kappa}\left(\delta v+\delta B\right)\label{Pert de curv comovel}
\end{equation}
which in Newtonian gauge is reduced to
\begin{equation}
\mathcal{R}=-\delta A+h_{\kappa}\delta v\label{Pert Curv New}
\end{equation}
where
\[
\delta v=-\frac{\delta Q}{Y-\sqrt{\frac{\beta_{0}e^{-3X}}{3}}T}
\]
with
\begin{equation}
\delta Q\equiv\delta X+\sqrt{\frac{\beta_{0}e^{-3X}}{3}}\delta W.\label{Var Muk Sas}
\end{equation}
In slow-roll leading order, $\delta A$ can be neglected in (\ref{Pert Curv New})
and the curvature perturbation is approximated by
\begin{equation}
\mathcal{R\simeq}-\frac{h_{\kappa}}{Y\left(1+\sqrt{\frac{\beta_{0}e^{X}}{3}}\right)}\delta Q,\label{Approx R}
\end{equation}
where $\delta Q$ is the generalization of Mukhanov-Sasaki variable\footnote{In the case of a single scalar field $\delta Q=\delta X$.}.
In addition, the denominator of Eq. (\ref{Approx R}) represents the
\emph{normalization} of $\delta v$ given by Eq.~(\ref{Normalization}).
It is important to stress that $\mathcal{R}$ is a gauge invariant
quantity which is conserved in super-horizon scales (see Appendix
\ref{sec - Conservacao R}). We also note that the normalization of
${\cal R}$ in Eq.~(\ref{Approx R}) is analogous to the two-field
inflation case \cite{Wand2008}; the difference being the non-canonical
kinetic factor $\beta_{0}e^{X}/3$ associated to the perturbation
$e^{-2X}\delta W$ \textemdash{} see Eqs.~(\ref{delta phi 1 aux})
and (\ref{delta phi 2 aux}).

A convenient combination of Eqs.~(\ref{Mu-Sa X}) and (\ref{Mu-Sa W})
leads to:
\begin{equation}
\delta\ddot{Q}+3h_{\kappa}\delta\dot{Q}-\bar{\nabla}^{2}\left(\delta Q\right)\simeq0.\label{Mu-Sa Q}
\end{equation}
By defining
\[
\delta\varphi\equiv aM_{Pl}\sqrt{\frac{3}{2}}\delta Q,
\]
it is possible to write Eq.~(\ref{Mu-Sa Q}) in Fourier space as
\begin{equation}
\delta\varphi_{_{\vec{k}}}^{\prime\prime}+\left(k^{2}-\frac{a^{\prime\prime}}{a}\right)\delta\varphi_{_{\vec{k}}}\simeq0.\label{Muk Sas padrao}
\end{equation}
This is the usual Mukhanov-Sasaki equation which can be quantized
in the standard way \cite{MuFeBran,BaumannLectures}. Thereby, the
dimensionless power spectrum $\Delta_{\delta Q}^{2}$ related to $\delta Q$
perturbation is given by
\begin{equation}
\Delta_{\delta Q}^{2}\simeq\frac{2}{3M_{Pl}^{2}}\left.\left(\frac{H}{2\pi}\right)^{2}\right\vert _{k=Ha}.\label{PSSH}
\end{equation}
The index $k=Ha$ indicates the power spectrum is calculated at the
specific time the perturbation crosses the horizon.

In order to compare the theoretical result with observations, it is
necessary to rewrite the power spectrum in terms of the curvature
perturbation. From Eqs.~(\ref{SR epsolon}), (\ref{Approx R}) and
(\ref{PSSH}), we obtain
\begin{equation}
\Delta_{\mathcal{R}}^{2}\simeq\frac{1}{8\pi^{2}M_{Pl}^{2}}\left.\frac{H^{2}}{\epsilon_{H}}\frac{\left(1-\sqrt{\frac{\beta_{0}e^{X}}{3}}\right)}{\left(1+\sqrt{\frac{\beta_{0}e^{X}}{3}}\right)}\right\vert _{k=Ha}.\label{Theoretical PS chi}
\end{equation}
This expression gives the curvature power spectrum in leading order
for the proposed inflationary model. The extra term with respect to
Starobinsky's action produces corrections in $\Delta_{\mathcal{R}}^{2}$
coming from both background dynamics (via the generalization of $\epsilon_{H}$)
and perturbations (through the generalization of $\mathcal{R}$).
In the next section, we will see how this extra term affects the predictions
of Starobinsky's inflation.


\subsection{Vector and tensor solutions}

During the inflationary regime, vector perturbations are described
by Eqs.~(\ref{vinculo vetorial}), (\ref{Eins Inv gauge vec 1}) and
(\ref{Eins Inv gauge vec 2}). By acting the operator $\delta^{jl}\partial_{l}$
onto (\ref{Eins Inv gauge vec 2}) then taking the Fourier transform,
we obtain
\[
-k^{2}\left(\delta\dot{F}_{i\left(\vec{k}\right)}+h_{\kappa}\delta F_{i\left(\vec{k}\right)}\right)=0,
\]
whose solution is
\[
\delta F_{i\left(\vec{k}\right)}=\frac{C_{i\left(\vec{k}\right)}}{a}.
\]
This shows $\delta F_{i}$ decays with $a\sim e^{Ht}$. As the other
two vector perturbations are identically null, Eq.~(\ref{all zero}),
we conclude that the proposed model does not generate any kind of
vector perturbation.

The expression~(\ref{Eins Inv gauge tensor}) associated to the
tensor perturbation $\delta E_{ij}$ is analogous to Eq.~(\ref{Mu-Sa Q}).
Decomposing $\delta E_{ij}$ as
\[
\delta E_{ij}=\frac{\delta v}{a}e_{ij},
\]
where $e_{ij}$ is the polarization tensor, and writing Eq.~(\ref{Eins Inv gauge tensor})
in Fourier space, results in:
\begin{equation}
\delta v_{_{\vec{k}}}^{\prime\prime}+\left(k^{2}-\frac{a^{\prime\prime}}{a}\right)\delta v_{_{\vec{k}}}=0.\label{Eq_pert_tens1}
\end{equation}
This is the standard Mukhanov-Sasaki equation for tensor perturbations.
Following an analogous approach to the scalar case \cite{LiddleLyth,MuFeBran},
one gets the tensor power spectrum:
\begin{equation}
\Delta_{\delta E_{ij}}^{2}\left(k\right)=\frac{8}{M_{Pl}^{2}}\left.\left(\frac{H}{2\pi}\right)^{2}\right\vert _{k=Ha}.\label{PSSH tensorial}
\end{equation}

It is worth mentioning $\delta E_{ij}$ is a gauge invariant quantity
which is conserved on super-horizon scales.


\section{Constraining the cosmological parameters\label{sec - Const Cos Par}}

The conservation of $\mathcal{R}$ and $\delta E_{ij}$ in super-horizon
scales allows to directly compare the inflationary power spectra Eqs.~(\ref{Theoretical PS chi}) and (\ref{PSSH tensorial}) with those
used as initial conditions in the description of CMB anisotropies.
This comparison is made through the parameterizations
\begin{align}
\Delta_{\mathcal{R}}^{2}\left(k\right) & =A_{s}\left(\frac{k}{k_{\ast}}\right)^{n_{s}-1},\label{Parametric PScalar}\\
\Delta_{h}^{2}\left(k\right) & =A_{t}\left(\frac{k}{k_{\ast}}\right)^{n_{t}},\label{Parametric PTensor}
\end{align}
where $A_{s}$ and $A_{t}$ are the scalar and tensor amplitudes,
$k_{\ast}$ is the pivot scale and $n_{s}$ and $n_{t}$ are the scalar
and tensor tilts \cite{Planck2018}.

By comparing Eqs.~(\ref{Theoretical PS chi}) and (\ref{PSSH tensorial})
to Eqs.~(\ref{Parametric PScalar}) and (\ref{Parametric PTensor}) and
using the slow-roll parameters $\epsilon_{H}$ and $\eta_{H}$ given
in Eqs.~(\ref{SR epsolon}) and (\ref{SR eta}), we obtain:
\begin{equation}
n_{s}\approx1+\eta_{H}-2\epsilon_{H}+\frac{4}{3}\frac{\sqrt{\frac{\beta_{0}e^{-X_{\ast}}}{3}}}{\left(1-\frac{\beta_{0}e^{X_{\ast}}}{3}\right)^{2}}\label{ns completo chii}
\end{equation}
and
\begin{equation}
r\equiv\frac{A_{t}}{A_{s}}\approx16\epsilon_{H}\frac{\left(1+\sqrt{\frac{\beta_{0}e^{X_{\ast}}}{3}}\right)}{\left(1-\sqrt{\frac{\beta_{0}e^{X_{\ast}}}{3}}\right)}.\label{r completo chii}
\end{equation}

In addition to $n_{s}$ and $r$, there is also
\begin{equation}
n_{t}\approx-2\epsilon_{H}\approx-\frac{r}{8}\frac{\left(1-\sqrt{\frac{\beta_{0}e^{X_{\ast}}}{3}}\right)}{\left(1+\sqrt{\frac{\beta_{0}e^{X_{\ast}}}{3}}\right)}.\label{nt completo chi}
\end{equation}
This expression shows how the consistency relation $n_{t}=-r/8$ \cite{LiddleLyth}
associated to a single scalar field inflation changes with the introduction
of the higher order term in Starobinsky action.

Eqs.~(\ref{ns completo chii}) and (\ref{r completo chii}) can be
written in terms of the $e$-folds number $N$, given by Eq.~(\ref{e-folds e X}).
Thus,
\begin{equation}
n_{s}\approx1+\frac{4\beta_{0}}{9}\frac{\left[1-3\sqrt{1-\frac{8}{9}\beta_{0}N_{\ast}}+\sqrt{1-\sqrt{1-\frac{8}{9}\beta_{0}N_{\ast}}}\right]}{\left(1-\sqrt{1-\frac{8}{9}\beta_{0}N_{\ast}}\right)\left(1-\frac{8}{9}\beta_{0}N_{\ast}\right)}\label{ns e-folds chi}
\end{equation}
and
\begin{equation}
r=\frac{32}{27}\frac{\beta_{0}^{2}}{\left(1-\sqrt{1-\frac{8}{9}\beta_{0}N_{\ast}}-\frac{4}{9}\beta_{0}N_{\ast}\right)\left(1-\sqrt{1-\sqrt{1-\frac{8}{9}\beta_{0}N_{\ast}}}\right)^{2}}\label{r e-folds chi}
\end{equation}
The results typical of Starobinsky inflation are recovered in the
limit $\beta_{0}\rightarrow0$:
\[
\lim_{\beta_{0}\rightarrow0}n_{s}\approx1-\frac{2}{N_{\ast}}\text{ \ and \ }\lim_{\beta_{0}\rightarrow0}r\approx\frac{12}{N_{\ast}^{2}}\text{.}
\]

\begin{figure}[h]
\includegraphics[scale=0.5]{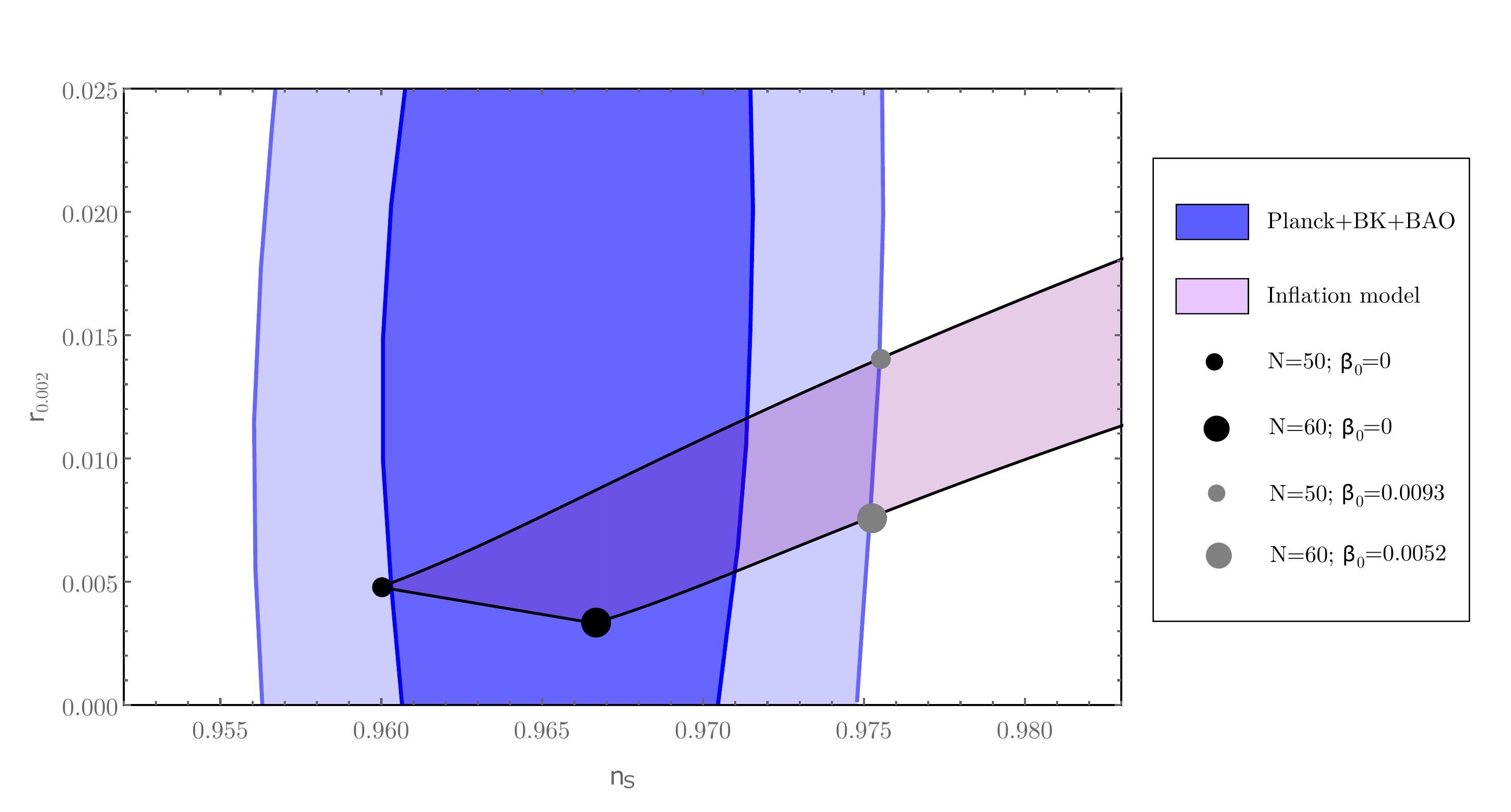}

\caption{The blue contours correspond to $68\%$ and $95\%$ CL constraints
on $n_{s}\times r$ given by Planck plus BICEP2/Keck plus BAO data
\cite{BK2018}. The black circles represent Starobinsky model
($\beta_{0}=0$). As $\beta_{0}$ increases the curves move to the
right (light purple region) increasing the tensor-to-scalar ratio
and the scalar tilt values. The grey circles represent the upper limits
for $\beta_{0}$ associated with $95\%$ CL for $n_{s}$. In this
case, $N_{\ast}=50$ corresponds to $\beta_{0}=9.3\times10^{-3}$
and $r=0.014$; and $N_{\ast}=60$ corresponds to $\beta_{0}=5.2\times10^{-3}$
and $r=0.0076$. \label{fig:X1X}}
\end{figure}

Fig. \ref{fig:X1X} displays the parametric plot $n_{s}\times r$ accounting
for the model with $\beta_{0} \neq 0$ and $50\leqslant N\leqslant60$.


Fig.~\ref{fig:X1X} shows how the addition of the term $\nabla^{\mu}R\nabla_{\mu}R$
in Starobinsky action increases the spectral index value and the
tensor-to-scalar ratio. The constraint of $95\%$ CL in $n_{s}$ sets
upper limits of $\beta_{0}=9.3\times10^{-3}$ and $\beta_{0}=5.2\times10^{-3}$
for $N_{\ast}=50$ and $N_{\ast}=60$, respectively. Thus, within
the observational limits, the proposed model is able to produce an
increase of up to $2.9$ times in the ratio $r$ when compared to
Starobinsky inflation. Furthermore, the value of $\beta_{0}\lesssim10^{-2}\ll1$
is consistent with the slow-roll approximation performed above. It
also guarantees a large range of initial conditions able to trigger
the inflationary regime (see Section \ref{InflationTermination}).

The energy density scale characteristic of inflation is determined
by $\varepsilon\approx\left(\kappa_{0}/4\right)M_{Pl}^{2}.$ From
Eqs. (\ref{Theoretical PS chi}) and (\ref{Parametric PScalar}) in
combination with Eq. (\ref{r e-folds chi}), we obtain
\[
\varepsilon\approx\frac{3\pi^{2}A_{s}}{2}rM_{Pl}^{4}\text{.}
\]
Moreover, we see in Fig.~\ref{fig:X1X} that the tensor-to-scalar
ratio varies from $0.0033\leq r\leq0.014$ within the range of $95\%$ CL.
Thus, for $A_{s}\simeq1.96\times10^{-9}$ \cite{Planck2018}, the energy density is in the range
\[
0.96\times10^{-10}\leq\frac{\varepsilon}{M_{Pl}^{4}}\leq4.06\times10^{-10},
\]
which is completely consistent with a sub-Planckian region.


\section{Discussion\label{sec:Discussion}}

In this work, we have studied the effects of a modification to
Starobinsky inflation model produced by the addition of the higher-order term
$\frac{\beta_{0}}{2\kappa_{0}^{2}}\nabla_{\mu}R\nabla^{\mu}R$. We
started from the generalized Starobinsky action in Einstein frame,
in which the Lagrangian depends on a scalar field $\Phi$ and an auxiliary
vector field $\xi^{\mu}$. We have constructed the cosmological background
dynamics and the perturbative structure of the theory for this model.

The background dynamics was determined from Friedmann equations as
well as from those for the auxiliary fields. After some manipulations,
we have shown the existence of an attractor region in the 4-dimensional
phase space ($\Phi,\dot{\Phi},\xi^{0},\dot{\xi}^{0}$) within $0<\beta_{0}<3/4$.
This attractor region is consistent with a slow-roll inflationary
period. In addition, we have seen that inflation ends with an oscillation
of the scalar field about the origin and with $\xi^{0}\rightarrow0$.
This characterizes an usual reheating phase.

The study of the perturbative regime was performed via the SVT decomposition,
as usual. We have shown that the vector and tensor degrees of freedom
behave just like in Starobinsky inflation. Also, we verified that
the curvature perturbation $\mathcal{R}$ (obtained from a proper
combination of the scalar degrees of freedom) satisfies Mukhanov-Sasaki
equation in the slow-roll leading order approximation. At last, we
obtained the tensorial and curvature power spectra and compared them
with the most recent observations from Keck Array and BICEP2 collaborations
\cite{BK2018}.

The main results from this work are summarized in Fig.~\ref{fig:X1X}
and indicate how the extra term $\frac{\beta_{0}}{2\kappa_{0}^{2}}\nabla_{\mu}R\nabla^{\mu}R$
changes the observable parameters of the primordial power spectrum.
In Fig. \ref{fig:X1X} we see that the parameter $\beta_{0}$ has
to be less than $10^{-2}$ for a number of e-folds in the interval
$50\leq N_{\ast}\leq60$. In this case, the scalar-to-tensor rate
$r$ can be expected to be up to three times the values predicted
by Starobinsky model. This result is particularly interesting since
it enables this natural generalization of Starobinsky model to
have an $r$ up to $\sim0.01$. Besides, the small values of $\beta_{0}$
preserves the chaotic structure of the theory, making room for a large
range of consistent initial conditions \cite{Berkin1990}.

The comparison of the results obtained here with
those of Ref.~\cite{Castellanos2018} shows some
interesting aspects. Firstly, it is important to realize that our
parameter $\beta_{0}$ is mapped on $-3k$ in Ref.~\cite{Castellanos2018}.
Hence, the results obtained here for $\beta_{0}>0$ have to be compared
to those obtained with $\gamma<0$ in Ref.~\cite{Castellanos2018}.
When comparing Fig.~4 of our paper with Fig.~4 in Ref.~\cite{Castellanos2018}, we observe that for very
low values of $\beta_{0}$ and $\gamma$ (red region in Ref.~\cite{Castellanos2018})
the models shade the same area. However, as these parameters are slightly
increased, discrepancies appear. We note that our model predicts higher
values for $r$ --- in a rough estimate, our values are about 3 times
greater than those of Ref.~\cite{Castellanos2018}.
Moreover, we realize that the values
of $-3k$ are at least three times greater than the equivalent values
of our $\beta_{0}$, when comparing the  68\% CL values for $\beta_{0}$ with the minimum values of $-3k$. These differences may be due to the
fact that the authors of Ref. \cite{Castellanos2018}
treated the higher order term as a small perturbation of the Starobinsky
model. This deserves a future and careful analyzis.

It must be highlighted that the action (\ref{Sg}) presents ghosts
for $\beta_{0}>0$ \cite{Hindawi1996}. On the other hand, the quantization
procedure performed in Section \ref{sec - sol. scalar fields} does
not show any pathology. The crucial point in this discussion is that scalar
perturbations in slow-roll leading order become constrained
by Eq.~(\ref{vinculo delta X e delta W}). Hence, there is only one
degree of freedom to be quantized. This degree of freedom is the (gauge
invariant) curvature perturbation, which is given by $\mathcal{R}\propto\delta Q=\delta X+\sqrt{\beta_{0}e^{-3X}/3}\delta W$
in the Newtonian gauge. Since the two terms composing $\delta Q$
have the same sign, we notice that $\delta Q$ is always a no-ghost
variable. As a consequence, it can be quantized as usual, independently
of the $\beta_{0}$ values. This situation is analogous to the treatment
given to ghosts by effective theories. Indeed, the energy scale remains
mostly unchanged during the quasi-exponential expansion (a fact that
is consistent with the slow-roll approximation) and the ghost degree
of freedom remains frozen.

The end of Section \ref{sec - Const Cos Par} shows that the energy scale
of the inflationary regime is sub-Planckian. This is a first indication
that the semi-classical approach adopted here is valid. The next step
is to check the naturalness, i.e. if the quantum corrections remain
small in this energy scale. This was addressed in \cite{Hertzberger2010}
for Starobinsky action\footnote{See also Ref. \cite{Copeland2015,Starobinsky2018} for the discussion in the context
of asymptotically safe gravity.} and, since $\beta_{0}<10^{-2}$, the result should apply to our model
as well. This subject shall be studied in a future work, where the
$\beta_{0}$ interval compatible with the requirement for naturalness
will be determined.

The higher-order modified Starobinsky inflation model presented here
can be further generalized to include the spin-$2$ terms $R_{\mu\nu}R^{\mu\nu}$
and $\nabla_{\mu}R_{\alpha\beta}\nabla^{\mu}R^{\alpha\beta}$ appearing
in action (\ref{Acao com dois Riemanns}). The effects upon inflation
coming from all these terms will be addressed by the authors in the
future.


\begin{acknowledgments}
The authors would like to thank Eduardo Messias de Morais for helping
with the figures. R.R. Cuzinatto acknowledges McGill University and
IFT-Unesp for hospitality and CAPES for partial financial support.
L.G. Medeiros acknowledges IFT-Unesp for hospitality and CNPq for
partial financial support. The authors would like
to thank the anonimous referee for the careful reading of the manuscript.
\end{acknowledgments}


\appendix

\section{Perturbation of $\delta\varepsilon$\label{sec - pert of energy density}}

The quantity $\delta\varepsilon$ is obtained by perturbing Eq.~(\ref{pressao FI})
and combining the result with Eq.~(\ref{Pert de epsilon mais p}).
After a long manipulation making use of the definitions (\ref{def 2})
and (\ref{def 3}), we obtain:
\begin{align}
\frac{2}{M_{Pl}^{2}\kappa_{0}}\delta\varepsilon & =-3Y^{2}\delta A+3Y\delta\dot{X}+e^{-X}\left(1-e^{-X}\right)\delta X\nonumber \\
 & +\beta_{0}e^{-X}\left(1-e^{-X}\right)\left[\delta\dot{T}+3h_{\kappa}\delta T-2Y\delta T-2T\delta\dot{X}\right]\nonumber \\
 & +\beta_{0}e^{-X}\left[\left(3h_{\kappa}T-2YT+S\right)\left(e^{-X}\delta F+2e^{-X}\left(1-e^{X}\right)\delta X\right)\right]\nonumber \\
 & -\beta_{0}^{2}e^{-2X}\left[\left(3h_{\kappa}T-2YT+S\right)\left[\left(3h_{\kappa}T-2YT+S\right)\delta X+T\left(\delta\dot{A}+\delta\dot{D}\right)+\bar{\nabla}^{2}\delta W\right]\right]\nonumber \\
 & +\beta_{0}e^{-2X}\left[\left(2YF-\dot{F}-3h_{\kappa}F\right)\delta T-F\delta\dot{T}+\left(2YT-3h_{\kappa}T-S\right)\delta F-T\delta\dot{F}\right]\nonumber \\
 & +2\beta_{0}e^{-2X}\left[\left(SF+T\dot{F}-2YTF+3h_{\kappa}TF\right)\delta X+2TF\delta\dot{X}\right]\nonumber \\
 & +3\beta_{0}e^{-3X}T\left[\frac{3}{2}T\delta X-\delta T-T\delta A\right],\label{Pert de epsilon}
\end{align}
where
\begin{align*}
F & =e^{X}-1+\beta_{0}\left(3h_{\kappa}T-2YT+S\right),\\
\delta F & =e^{X}\delta X+\beta_{0}\left[T\delta\dot{D}+\delta\dot{T}+3h_{\kappa}\delta T+\bar{\nabla}^{2}\delta W-2Y\delta T-2T\delta\dot{X}\right],
\end{align*}
and
\[
\delta D=-\frac{\delta g}{2a^{6}}=\delta A+3\delta C+a^{2}\bar{\nabla}^{2}\delta E,
\]
where $\delta g$ is the perturbation in the metric determinant.

\section{Conservation of comoving curvature perturbation\label{sec - Conservacao R}}

The first step to show that $\mathcal{R}$ is conserved in super-horizon
scales is to determine $\mathcal{\dot{R}}$. We derive $\mathcal{R}$
as given by (\ref{Pert Curv New}) and use the equations on the background
\textemdash{} (\ref{Friedmann one1}) and (\ref{Friedmann two2})
\textemdash{} and the equations of the perturbative part \textemdash{}
(\ref{Eins Inv gauge esc 1}), (\ref{Eins Inv gauge esc 2}) and (\ref{Eins Inv gauge esc 3}).
In this way,
\begin{equation}
\mathcal{\dot{R}}=-\frac{h_{\kappa}}{\left(\varepsilon+p\right)}\left[\delta p_{\text{nad}}-2\kappa_{0}M_{Pl}^{2}\frac{\dot{p}}{\dot{\varepsilon}}\bar{\nabla}^{2}\Psi\right]-\bar{\partial}_{0}\left(\frac{h_{\kappa}\delta q}{\left(\varepsilon+p\right)}\right),\label{Var de pert de curvatura}
\end{equation}
where
\[
\delta p_{\text{nad}}\equiv\delta p-\frac{\dot{p}}{\dot{\varepsilon}}\delta\varepsilon\text{ \ and \ }\delta A=-\Psi.
\]

The next step is to show that $\mathcal{\dot{R}}\propto\nabla^{2}\Psi$
during the inflationary regime. We start by manipulating Einstein
equations (\ref{Eins Inv gauge esc 1}) and (\ref{Eins Inv gauge esc 2})
to obtain
\begin{equation}
\delta\varepsilon=-2\kappa_{0}M_{Pl}^{2}\bar{\nabla}^{2}\Psi+3h_{\kappa}\left[\left(\varepsilon+p\right)\delta v+\delta q\right].\label{pert eps combinado}
\end{equation}
Then, we substitute the conservation equation $\dot{\varepsilon}=3h_{\kappa}\left(\varepsilon+p\right)$ and Eq.~(\ref{pert eps combinado}) into (\ref{Var de pert de curvatura}):
\[
\mathcal{\dot{R}}=-\frac{h_{\kappa}}{\left(\varepsilon+p\right)}\left[\delta\varepsilon+\delta p+\left(\dot{\varepsilon}+\dot{p}\right)\delta v+2\kappa_{0}M_{Pl}^{2}\bar{\nabla}^{2}\Psi+\delta\dot{q}+\frac{\dot{h}_{\kappa}}{h_{\kappa}}\delta q\right].
\]
In leading order of slow-roll, the terms of the background are classified
as
\[
h_{\kappa}\sim1,\text{ \ }\dot{h}_{\kappa}\sim e^{-2X}\text{ and }\dot{\varepsilon}+\dot{p}\sim e^{-3X}.
\]
Moreover, the perturbative quantities in Newtonian gauge are approximated
by:
\begin{align*}
\delta\varepsilon+\delta p & \simeq3\kappa_{0}M_{Pl}^{2}Y\left[\delta\dot{X}+\frac{\beta_{0}}{3}e^{-X}\delta\dot{W}\right]\Rightarrow\delta\varepsilon+\delta p\sim e^{-X}\delta X,\\
\delta v & \simeq-\frac{\delta X+\sqrt{\frac{\beta_{0}e^{-3X}}{3}}\delta W}{Y\left(1+\sqrt{\frac{\beta_{0}e^{X}}{3}}\right)}\Rightarrow\delta v\sim e^{X}\delta X,\\
\delta q & \simeq-\frac{3}{2}\sqrt{\frac{\beta_{0}e^{X}}{3}}\kappa_{0}M_{Pl}^{2}Y\left(\delta X-e^{-2X}\delta W\right)\Rightarrow\delta q\sim e^{-2X}\delta X,
\end{align*}
where the slow-roll approximations were used together with the relations
(\ref{consis 1}) and (\ref{consis 2}). Note that $\delta q$ is
suppressed by an extra order in slow-roll due to the tracking solution
(\ref{vinculo delta X e delta W}). Thus, up to leading order, $\mathcal{\dot{R}}$
is approximated by
\begin{equation}
\mathcal{\dot{R}}\simeq-\frac{3}{2}\kappa_{0}M_{Pl}^{2}\frac{h_{\kappa}}{\left(\varepsilon+p\right)}\left[2Y\left(\delta\dot{X}-\frac{\beta_{0}}{3}e^{-X}\delta\dot{W}\right)+\frac{4}{3}\bar{\nabla}^{2}\Psi\right].\label{Aux R cali}
\end{equation}

The following step is to write the quantity $\left( \delta\dot{X}-\frac{\beta_{0}}{3}e^{-X}\delta\dot{W} \right)$
in a more convenient form. By approximating Eq.~(\ref{Pert de epsilon})
up to leading order and using the Eqs.~(\ref{Mu-Sa W}) and (\ref{vinculo delta X e delta W}),
one obtains, after a long manipulation:
\begin{equation}
\frac{2}{3M_{Pl}^{2}\kappa_{0}}\delta\varepsilon\simeq Y\left[\delta\dot{X}-\frac{\beta_{0}}{3}e^{-X}\delta\dot{W}\right]+\frac{1}{3}e^{-X}\delta X.\label{Aux delta eps}
\end{equation}
On the other hand, in the attractor region (\ref{Phi adm aprox final}),
\[
e^{-X}\simeq-9Yh_{\kappa}\left(1-\frac{\beta_{0}}{3}e^{X}\right).
\]
Thus, Eq. (\ref{Aux delta eps}) is cast in the form:
\[
\frac{2}{3M_{Pl}^{2}\kappa_{0}}\delta\varepsilon\simeq Y\left[\delta\dot{X}-\frac{\beta_{0}}{3}e^{-X}\delta\dot{W}-3h_{\kappa}\left(1-\frac{\beta_{0}}{3}e^{X}\right)\delta X\right].
\]
Comparison with Eq.~(\ref{pert eps combinado}) leads to:
\begin{equation}
\frac{2}{3M_{Pl}^{2}\kappa_{0}}\left[\left(\varepsilon+p\right)\delta v+\delta q\right]\simeq\frac{1}{3h_{\kappa}}\left[Y\left(\delta\dot{X}-\frac{\beta_{0}}{3}e^{-X}\delta\dot{W}\right)-3Yh_{\kappa}\left(1-\frac{\beta_{0}}{3}e^{X}\right)\delta X+\frac{4}{3}\bar{\nabla}^{2}\Psi\right].\label{Aux Appendix}
\end{equation}
In addition, using the approximation $T\simeq-Ye^{2X}$ and Eqs.~(\ref{energy plus pressure}),
(\ref{Pert v}), (\ref{Pert q}) and (\ref{vinculo delta X e delta W}),
one can write the left side of (\ref{Aux Appendix}) as:
\begin{equation}
\frac{2}{3M_{Pl}^{2}\kappa_{0}}\left[\left(\varepsilon+p\right)\delta v+\delta q\right]\simeq-Y\left(\delta X-\frac{\beta_{0}}{3}e^{-X}\delta W\right)\simeq-Y\left(1-\frac{\beta_{0}}{3}e^{X}\right)\delta X.\label{Aux appendix 2}
\end{equation}
Therefore, the Eq.~(\ref{Aux Appendix}) is simplified to
\begin{equation}
Y\left(\delta\dot{X}-\frac{\beta_{0}}{3}e^{-X}\delta\dot{W}\right)\simeq-\frac{4}{3}\bar{\nabla}^{2}\Psi.\label{Aux appendix 1}
\end{equation}
Finally, Eq.~(\ref{Aux appendix 1}) is replaced into (\ref{Aux R cali})
so that $\mathcal{\dot{R}}$ assumes the form:
\begin{equation}
\mathcal{\dot{R}}\simeq\frac{2\kappa_{0}M_{Pl}^{2}}{\left(\varepsilon+p\right)}h_{\kappa}\bar{\nabla}^{2}\Psi.\label{R conservado}
\end{equation}

Eq.~(\ref{R conservado}) can be rewritten as
\[
\frac{d\mathcal{R}}{d\ln a}\sim\left(\frac{k}{aH}\right)^{2}\Psi.
\]
Hence, $\mathcal{R}$ is conserved in super-horizon scales ($k\ll aH$)
in slow-roll leading order. Moreover, toward the end of inflation,
where $e^{-X}\sim1$, the vector field $\xi^{\mu}$ becomes negligible
by a factor $\sqrt{\beta_{0}}$. As the observational limits impose
$\beta_{0}\lesssim10^{-2}$ cf. Section \ref{sec - Const Cos Par}),
the end of inflation occurs similarly to the case of a single scalar
field (see Fig.~\ref{fig:XY}). Thus, assuming that during reheating
and the entire hot universe the non-adiabatic perturbations $\delta p_{\text{nad}}$
are negligible, we conclude from (\ref{Var de pert de curvatura})
that $\mathcal{R}$ generated in inflation remains (approximately)
constant throughout its super-horizon evolution.



\end{document}